\documentclass[11pt]{article}

\usepackage{fullpage}
\usepackage{amsthm}
\usepackage{amsfonts}
\usepackage[colorlinks=false]{hyperref}

\newcommand{\shortitem}{\vspace{-.25cm}\item}

\newcommand{\ceil}[1]{{\left \lceil #1 \right \rceil}}
\newcommand{\Hmin}{H_{\infty}}
\newcommand{\Adv}{\mathbf{Adv}}
\newcommand{\InSec}{\mathbf{InSec}}
\newcommand{\Sec}{\mathbf{Sec}}
\newcommand{\Rel}{\mathbf{Rel}}
\newcommand{\UnRel}{\mathbf{UnRel}}
\newcommand{\stl}{\mathrm{STL}}
\newcommand{\stf}{\mathrm{STF}}
\newcommand{\st}{\mathit{ST}}
\newcommand{\CC}{\mathcal{C}}
\newcommand{\his}{\mathcal{H}}
\newcommand{\SE}{\mathit{SE}}
\newcommand{\SD}{\mathit{SD}}
\newcommand{\qs}{\mathcal{S}}
\newcommand{\F}{\mathcal{F}}
\newcommand{\seed}{\mathrm{seed}}
\newcommand{\outflag}{\mathit{f}}
\newcommand{\ctr}{\mathit{ctr}}
\newcommand{\eps}{\varepsilon}

\newcommand{\chanRate}{w}  
\newcommand{\natE}{e}      
\newcommand{\RScountk}{k}  
\newcommand{\key}{K}       

\newcommand{\maxprob}{p}

\newcommand{\kwfont}[1]{{\texttt{#1}}}

\newcommand{\kwfor}{{\kwfont{for}\ }}
\newcommand{\kwto}{{\kwfont{to}\ }}

\newcommand{\kwor}{{\kwfont{or}\ }}

\newcommand{\kwuntil}{{\kwfont{until}\ }}
\newcommand{\kwrepeat}{{\kwfont{repeat}\ }}
\newcommand{\kwif}{{\kwfont{if}\ }}
\newcommand{\kwthen}{{\kwfont{then}\ }}
\newcommand{\kwelse}{{\kwfont{else}\ }}

\newcommand{\kwabort}{{\kwfont{abort}\ }}

\newcommand{\kwoutput}{{\kwfont{output}\ }}
\newcommand{\kwlet}{{\kwfont{let}\ }}

\def\bool{\{0, 1\}}

\newcommand{\leftrand}{\buildrel{R}\over{\leftarrow}}
\def\pr{\Pr}
\newcommand\expc[1]{\mathrm{E}[#1]}

\newcommand{\nat}{\mathbb{N}}

\newcommand{\suchthat}{\; | \;}
\newcommand{\aand}{\; \wedge \;}

\newcommand{\emdefinition}[2]
	   {
	     \noindent
	     \begin{tabular}{|l|l|}
	       \begin{minipage}{5cm} &
	       \begin{minipage}{10cm}
		 \setlength{\parindent}{-0.1\textwidth}
		 \par \begin{bf}Definition (#1)\end{bf} #2
	       \end{minipage}
	     \end{tabular}
	   }

\newcommand{\rndelt}{\mathtt{rndelt}}
\newcommand{\insupp}{\mathtt{insupp}}

\newcommand{\epr}{\mathit{EPR}}
\newcommand{\dpr}{\mathit{DPR}}
\newcommand{\inspr}{\ensuremath{\mathit{InsPR}}}
\newcommand{\nqpr}{\ensuremath{\mathit{NqPR}}}
\newcommand{\corrpr}{\ensuremath{\mathit{CorrPR}}}
\newcommand{\fewpr}{\ensuremath{\mathit{FewPR}}}
\newcommand{\lowpr}{\ensuremath{\mathit{LowPR}}}
\newcommand{\nq}{\ensuremath{\mathit{Nq}}}
\newcommand{\ins}{\ensuremath{\mathit{Ins}}}
\newcommand{\corr}{\ensuremath{\mathit{Corr}}}
\newcommand{\few}{\ensuremath{\mathit{Few}}}
\newcommand{\low}{\ensuremath{\mathit{Low}}}
\newcommand{\iprf}{\ensuremath{\iota_{PRF}}}

\newcommand{\addition}[1]{#1}

\newtheorem{lemma}{Lemma}
\newtheorem{proposition}{Proposition}
\newtheorem{claim}{Claim}
\newtheorem{theorem}{Theorem}
\theoremstyle{definition}
\newtheorem{definition}{Definition}

\date{March 31, 2008}
\title{Upper and Lower Bounds on Black-Box
Steganography\footnote{Preliminary version appears in TCC 2005 \cite{DIRR05}.}}
\author{Nenad Dedi\'c \and Gene Itkis \and Leonid Reyzin \and Scott Russell \and\\
Boston University\\Department of Computer Science\\111 Cummington
Street\\Boston, MA 02215\\ \texttt{\small\{nenad,itkis,reyzin,srussell\}@cs.bu.edu}}

\begin{document}
\maketitle

\begin{abstract}
We study the limitations of steganography when the sender is not using any
properties of the underlying channel beyond its entropy and the ability to
sample from it.  On the negative side, we show that the number of samples
the sender must obtain from the channel is exponential in the rate of the
stegosystem.  On the positive side, we present the first secret-key
stegosystem that essentially matches this lower bound regardless of the
entropy of the underlying channel.  Furthermore, for high-entropy channels,
we present the first secret-key stegosystem that matches this lower bound
{\em statelessly} (i.e., without requiring synchronized state between
sender and receiver).

\medskip

\end{abstract}

\section{Introduction}

Steganography's goal is to conceal the presence of a secret message within
an innocuous-looking communication.  In other words, steganography consists
of hiding a secret {\em hiddentext} message within a public {\em covertext}
to obtain a {\em stegotext} in such a way that an unauthorized observer 
is unable to distinguish between a
covertext {\em with} a hiddentext and one {\em without}.

The first rigorous complexity-theoretic formulation of secret-key
steganography was provided by Hopper, Langford and von Ahn
\cite{HLvA02cor}.  In this formulation, \emph{steganographic secrecy} of
a stegosystem is defined as the inability of a polynomial-time
adversary to distinguish between observed distributions of unaltered
covertexts and stegotexts.  (This is in contrast with many previous
works, which tended to be information-theoretic in perspective; see,
e.g., \cite{Cac98} and other references in \cite{HLvA02cor,Cac98}.)

\subsection{Model}
In steganography, the very presence of a message must be hidden from the
adversary, who must be given no reason for suspecting that anything is
unusual.  This is the main difference from encryption, which does not
prevent the adversary from suspecting that a secret message is being sent,
but only from decoding the message.  To formalize ``unusual,''
some notion of usual communication must exist.

We adopt the model of \cite{HLvA02cor} with minor changes.  In it, 
\emph{sender} sends data to \emph{receiver}.  The usual (nonsteganographic) communication comes from
the \emph{channel}, which is a distribution of possible \emph{documents}
sent from sender to receiver based on past communication.  The channel
models the sender's decision process about what to say next in ordinary
communication; thus, the sender is given access to the channel via a {\em
sampling oracle} that takes the past communication as input and returns the
next document from the appropriate probability distribution.  Sender and
receiver share a secret key (public-key steganography is addressed in
\cite{vAH04,BC05}).

The adversary is assumed to also have some information about the usual
communication, and thus about the channel.  It listens to the communication
and tries to distinguish the case where the sender and receiver are just
carrying on the usual conversation (equivalently, sender is honestly
sampling from the oracle) from the case where the sender is transmitting a
hiddentext message $m\in\bool^*$ (the message may even be chosen by the
adversary).  A stegosystem is secure if the adversary's suspicion is not
aroused---i.e., if the two cases cannot be distinguished.

\subsection{Desirable Characteristics of a Stegosystem}
\paragraph*{Black-Box.}
In order to obtain a stegosystem of broad applicability, one would like to
make as few assumptions as possible about the understanding of the
underlying channel.  As Hopper et al. \cite{HLvA02cor} point out, the
channel may be very complex and not easily described.  For example, if the
parties are using photographs of city scenes as covertexts, it is
reasonable to assume that the sender can obtain such photographs, but
unreasonable to expect the sender and the receiver to know a
polynomial-time algorithm that can construct such photographs from
uniformly distributed random strings.  We therefore concentrate on {\em
black-box} steganography, in which the knowledge about the channel is
limited to the sender's ability to query the sampling oracle and a bound
on the channel's min-entropy available to sender and receiver.  In
particular, the receiver is not assumed to be able to sample from the channel.
The adversary, of course, may know more about the channel.

\paragraph*{Efficient (in terms of running time, number of samples, rate, reliability).}
The running times of sender's and receiver's
algorithms should be minimized.  Affairs are slightly complicated by
the sender's algorithm, which involves two kinds of
fundamentally different operations: \emph{computation}, and \emph{channel sampling}.  Because obtaining a channel sample could conceivably be of much higher cost
than performing a computation step, the two should be separately accounted for.

{\em Transmission rate} of a stegosystem is the number of hiddentext bits
transmitted per single stegotext document sent.  
Transmission rate is tied to {\em reliability}, which is the
probability of successful decoding of an encoded message (and {\em
unreliability}, which is one minus reliability).  The goal is to
construct stegosystems that are reliable and transmit at a high rate (it is
easier to transmit at a high rate if reliability is low and so the
receiver will not understand much of what is transmitted).

Even if a stegosystem is black-box, its efficiency may depend
on the channel distribution.  We will be interested in the dependence on the
channel min-entropy $h$.  Ideally, a stegosystem would work well even for
low-min-entropy channels.

\paragraph*{Secure.}
\emph{Insecurity} is defined as the adversary's advantage in
distinguishing stegotext from regular channel communication  (and {\em security} as one minus
insecurity).  Note that security, like efficiency, may depend on the channel
min-entropy. We are interested in stegosystems with insecurity
as close to 0 as possible, ideally even for low-min-entropy channels.

\paragraph*{Stateless.}
It is desirable to construct {\em stateless} stegosystems, so that the
sender and the receiver need not maintain synchronized state in order to
communicate long messages.  Indeed, the need for
synchrony may present a particular problem in steganography in case
messages between sender and receiver are dropped or arrive out of order.
Unlike in counter-mode symmetric encryption, where the counter value can be
sent along with the ciphertext in the clear, here this is not possible: the
counter itself would also have to be steganographically encoded to avoid
detection, which brings us back to the original problem of
steganographically encoding multibit messages.

\subsection{Our Contributions}
We study the optimal efficiency achievable by black-box steganography, and
present secret-key stegosystems that are nearly optimal.
Specifically, we demonstrate the following results:
\begin{itemize}
\item A lower bound, which states that
a secure and reliable black-box stegosystem with rate of $w$ bits per document
sent requires the encoder to take at least $c 2^w$ samples from the channel
per $w$ bits sent, for some constant $c$.  The value of $c$ depends
on security and reliability, and tends to $1/(2e)$ as security and reliability
approach 1.  This lower bound applies to secret-key as well as
public-key stegosystems.

\item A stateful black-box secret-key stegosystem $\stf$ that transmits $w$
bits per document sent, takes $2^w$ samples per $w$ bits, and has unreliability
of $2^{-h+w}$ per document (recall that $h$ is the channel entropy)
and negligible insecurity, which is independent
of the channel. \addition{(A very similar construction
was independently discovered by Hopper~\cite[Construction 6.10]{Hop04}.)
}

\item A stateless black-box secret-key stegosystem $\stl$  that
transmits $w$ bits per document sent, takes $2^w$ samples per $w$ bits, and
has unreliability $2^{-\Theta(2^{h})}$ and insecurity negligibly
close to $l^2 2^{-h+2w}$ for $lw$ bits sent.
\end{itemize}

\noindent
Note that for both stegosystems, the rate vs. number of samples tradeoff is
very close to the lower bound---in fact, for channels with sufficient
entropy, the optimal rate allowed by the lower bound and the achieved rate
differ by $\log_2 2e < 2.5$ bits (and some of that seems due to slack in
the bound).  Thus, our bound is quite tight, and our stegosystems quite
efficient.  The proof of the lower bound involves a surprising application
of the huge random objects of \cite{GGN03}, specifically of the truthful
implementation of a boolean function with interval-sum queries.  The
lower bound demonstrates that significant improvements in stegosystem
performance must come from assumptions about the channel.

The stateless stegosystem $\stl$ can be used whenever the underlying channel
distribution has sufficient min-entropy $h$ for the insecurity $l^2 2^{-h+2w}$ to be
acceptably low. It is extremely simple,
requiring just evaluations of a pseudorandom function for encoding and
decoding, and very reliable.

If the underlying channel does not have sufficient min-entropy, then the
stateful stegosystem $\stf$ can be used, because its insecurity is
independent of the channel.  While it requires shared synchronized state
between sender and receiver, the state information is only a counter of the
number of documents sent so far. If min-entropy of the channel is so low
that unreliability of $2^{-h+w}$ per document is too high for the application,
reliability of this stegosystem can be improved through the use of
error-correcting codes over the $2^{w}$-ary alphabet (applied to the
hiddentext before stegoencoding), because failure to decode correctly is
independent for each $w$-bit block.  Error-correcting codes can increase
reliability to be negligibly close to 1 at the expense of reducing the
asymptotic rate from $w$ to $w-(h+2)2^{-h+w}$.  Finally, of course, the
min-entropy of any channel can be improved from $h$ to $nh$ by viewing $n$
consecutive samples as a single draw from the channel; if $h$ is extremely
small to begin with, this will be more efficient than using
error-correcting codes (this improvement requires both parties to be
synchronized modulo $n$, which is not a problem in the stateful case).

This stateful stegosystem $\stf$ also admits a few variants.  First, the
logarithmic amount of shared state can be eliminated at the expense of
adding a linear amount of private state to the sender and reducing
reliability slightly (as further described in~\ref{section-stf-stateless}), thus
removing the need for synchronization between the sender and the receiver.
Second, under additional assumptions about the channel (e.g., if each
document includes time sent, or has a sequence number), $\stf$  can be made
completely stateless. \addition{The remarks of this paragraph and the
  previous one can be equally applied to~\cite[Construction 6.10]{Hop04}.}

\subsection{Related Work}
The bibliography on the subject of steganography is extensive; we do not review
it all here, but rather recommend references in \cite{HLvA02cor}.

\paragraph{Constructions.}
In addition to introducing the complexity-theoretic model for
steganography, \cite{HLvA02cor} proposed two constructions of black-box\footnote{%
Construction 2, which, strictly
speaking, is not presented as a black-box construction in \cite{HLvA02cor},
can be made black-box through the use of extractors (such as universal hash
functions) in place of unbiased functions, as shown in~\cite{vAH04}.}
secret-key stegosystems, called Construction 1 and Construction
2.

Construction 1 is stateful and, like our stateful construction $\stf$, boasts
negligible insecurity regardless of the channel.  However, it can transmit
only 1 bit per document, and its reliability is limited by
$1/2+1/4(1-2^{-h})$ per document sent, which means that, regardless of the
channel, each hiddentext bit has probability at least $1/4$ of arriving
incorrectly (thus, to achieve high reliability, error-correcting codes with
expansion factor of at least $1/(1-H_2(1/4))\approx 5$ are needed).  In
contrast, $\stf$
has reliability that is exponentially (in the min-entropy) close to 1, and
thus works well for any channel with sufficient entropy.  Furthermore, it
can transmit at rate $w$ for any $w<h$, provided that  the encoder has sufficient
time for the $2^w$ samples required.  It can be seen as a generalization of
Construction 1.

Construction 2 of~\cite{HLvA02cor} is stateless.  Like the security of our
stateless construction $\stl$, its security depends on the min-entropy of the
underlying channel.  While no exact analysis is provided
in~\cite{HLvA02cor}, the insecurity of Construction 2 seems to be roughly
$\sqrt{l}2^{(-h+w)/2}$ (due to the fact that the adversary sees $l$ samples
either from $\CC$ or from a known distribution with bias
roughly $2^{(-h+w)/2}$ caused by a public extractor; see Appendix~\ref{appendix-eps-biased}), which is
higher than the insecurity of $\stl$  (unless $l$ and
$w$ are so high that $h<3w+3\log l$, in which case both constructions are
essentially insecure, because insecurity is higher than the inverse of the
encoder's running time $l2^w$).  Reliability of Construction 2, while not
analyzed in \cite{HLvA02cor}, seems close to the reliability of $\stl$.
The rate of Construction 2 is lower (if other parameters
are kept the same), due to the need for randomized encryption of the
hiddentext, which necessarily expands the number of bits sent.

It is important to note that the novelty of $\stl$ is
not the construction itself, but rather its analysis.  Specifically, its
stateful variant appeared as Construction 1 in the Extended Abstract of
\cite{HLvA02cor}, but the analysis of the Extended Abstract 
was later found to be flawed by
\cite{Lea-Tal-Omer}.  Thus, the full version of \cite{HLvA02cor} included a
different Construction 1.  We simply revive this old construction, make it
stateless, generalize it to $w$ bits per document, and, most importantly,
provide a new analysis for it.

In addition to the two constructions of~\cite{HLvA02cor} described above,
and independently of our work, Hopper~\cite{Hop04} proposed two more
constructions: Constructions 6.10 ({\tt MultiBlock}) and 3.15 ({\tt
NoState}).  As already mentioned, {\tt MultiBlock} is essentially the same
as our $\stf$.   {\tt NoState} is an
interesting variation of Construction 1 of \cite{HLvA02cor} that addresses
the problem of maintaining shared state at the expense of lowering the rate
even further.

\paragraph{Bounds on the Rate and Efficiency.}
\addition{Hopper in \cite[Section 6.2]{Hop04} establishes a bound on the
 rate vs. efficiency tradeoff.  Though quantitatively similar to ours (in
 fact, tighter by the constant of $2e$), this bound applies only to a
 restricted class of black-box stegosystems: essentially, stegosystems that
 encode and decode one block at a time and sample a fixed number of
 documents per block.  The bound presented in this paper applies to any
 black-box stegosystem, as long as it works for a certain reasonable class
 of channels, and thus can be seen as a generalization of the bound
 of~\cite{Hop04}.  Our proof techniques are quite different than those
 of~\cite{Hop04}, and we hope they may be of independent interest.  We
 refer the reader to Section \ref{section-bounded-adversaries} for an
 elaboration.
  Finally it should be noted that non-black-box
  stegosystems can be much more efficient---see
  \cite{HLvA02cor,vAH04,Le03,Le-Kurosawa}.}

\section{Definitions}
\label{section-definitions}
\subsection{Steganography}
The definitions here are essentially those of \cite{HLvA02cor}.  We modify
them in three ways.  First, we view the channel as producing
documents (symbols in some, possibly very large, alphabet) rather than
bits.  This simplifies notation and makes min-entropy of the channel more
explicit.  Second, we consider stegosystem reliability as a parameter rather
than a fixed value.  Third, we make the length of the adversary's
description (and the adversary's dependence on the channel) explicit
in the definition.

\paragraph*{The Channel.}
Let $\Sigma$ be an alphabet; we call the elements of $\Sigma$
\emph{documents}. A channel $\CC$ is a map that takes a history
$\his\in \Sigma^*$ as input and produces a probability distribution $D_\his\in
\Sigma$.  A history $\his= s_1 s_2 ... s_n$ is {\em legal}  if each
subsequent symbol is obtainable given the previous ones, i.e.,
$Pr_{D_{s_1 s_2 \dots s_{i-1}}}[s_i]>0$.
Min-entropy of a distribution $D$ is defined as $\Hmin(D) =
\min_{s\in D} \{-\log_2 \Pr_D[s]\}$.  Min-entropy of $\CC$ is the $\min_\his
\Hmin(D_\his)$, where the minimum is taken over legal histories $\his$.

Our stegosystems will make use of a channel sampling oracle $M$, which, on
input $\his$, outputs a symbol $s$ according to $D_\his$.  A stegosystem
may be designed for a particular $\Sigma$ and min-entropy of $\CC$.

\begin{definition}\label{def-stegosystem}
A \emph{black-box secret-key stegosystem} for the alphabet $\Sigma$ is a pair of
probabilistic polynomial time algorithms $\st=(\SE,\SD)$ such that, for a
security parameter $\kappa$,
\begin{enumerate}
\item $\SE$ has access to a channel sampling oracle $M$ for a channel $\CC$
on $\Sigma$ and
  takes as input a randomly chosen key $K\in \{0,1\}^\kappa$,
  a string
  $m\in \{0,1\}^*$ (called the \emph{hiddentext}), and the channel history
$\his$.
  It returns a string
  of symbols $s_1 s_2\dots s_l \in \Sigma^*$ (called the \emph{stegotext})
                                                                                
\item $\SD$ takes as input a key $K\in \{0,1\}^\kappa$, a stegotext
$s_1 s_2\dots s_l \in \Sigma^*$, and a channel history $\his$
and returns a hiddentext $m\in \{0,1\}^*$.
\end{enumerate}
\noindent We further assume that the length $l$ of the stegotext output by
$\SE$ depends only on the length of hiddentext $m$ but not on its contents.
\end{definition}

\paragraph*{Stegosystem Reliability.}
The \emph{reliability} of a
stegosystem $\st$ with security parameter $\kappa$
for a channel $\CC$ and messages of length $\mu$ is defined as
\[
\Rel_{\st(\kappa),\CC,\mu}= \min_{m\in \{0,1\}^\mu,\his}
\{\Pr_{K\in\{0,1\}^\kappa} [\SD(K,SE^M(K,m, \his), \his)=m]\}\,.
\]
Unreliability is defined as
$\UnRel_{\st(\kappa), \CC, \mu} = 1-\Rel_{\st(\kappa), \CC, \mu}$.

\paragraph*{The Adversary.}
We consider only passive adversaries who mount
a chosen hiddentext attack on $\st$ (stronger adversarial models
for steganography have also been considered, see, e.g., \cite{HLvA02cor,vAH04,BC05}).
The goal of such an adversary is to distinguish whether it is seeing
encodings of the hiddentext it supplied to the encoder or simply random
draws from the channel.  To this end, define an oracle $O(\cdot, \his)$ that
produces random draws from the channel starting with history $\his$ as
follows: on input $m\in \{0,1\}^*$, $O$ computes the length $l$
of the stegotext that $\SE^M(K,m)$ would have output and outputs
$s_1 s_2 \dots s_l$ where each $s_i$ is drawn according to
$D_{\his \circ s_1 s_2 \dots s_{i-1}}$.

\begin{definition} $W$
is a \emph{$(t,d,q,\lambda)$ passive adversary for stegosystem $\st$} if
\begin{enumerate}
\shortitem $W$ runs in expected time $t$ (including the running time needed by the
stegoencoder to answer its queries) and has description of length $d$ (in some
canonical language).
\shortitem $W$ has access to $\CC$ via the sampling oracle $M(\cdot)$.
\shortitem $W$ can make an expected number of $q$ queries of combined length $\lambda$ bits to an
  oracle
  which is either $\SE^M(K,\cdot, \cdot)$ or $O(\cdot, \cdot)$.
\shortitem $W$ outputs a bit indicating whether it was interacting with $\SE$ or
  with $O$.
\end{enumerate}
\end{definition}

\paragraph{Stegosystem Security.}
The {\em advantage} $\Adv^\mathrm{SS}$ (here $\mathrm{SS}$ stands for
``Steganographic Secrecy'') of
$W$ against $\st$ with security parameter $\kappa$
for a channel $\CC$ is 
defined as
\[
\Adv^\mathrm{SS}_{\st(\kappa),\CC}(W)=
\left|\Pr_{K\gets \{0,1\}^\kappa}[W^{M,\SE^M(K,\cdot,\cdot)}=1]-
   \Pr[W^{M,O(\cdot,\cdot)}=1]\right|\,.
\]
For a given $(t, d, q, \lambda)$, the {\em insecurity} of a stegosystem $\st$ with
respect to channel $\CC$ is defined as
\[\InSec^\mathrm{SS}_{\st(\kappa),\CC}(t,d,q,\lambda)=
   \max_{(t,d,q,\lambda) \mbox{ adversary } W} \{{\bf
     Adv}^\mathrm{SS}_{\st(\kappa),\CC}(W)\}\,,
\]
and security $\Sec$ as $1-\InSec$.

Note that the adversary's algorithm can depend on the channel $\CC$,
subject to the restriction on the algorithm's total length $d$.  In other
words, the adversary can possess some description of the channel in
addition to the black-box access provided by
the channel oracle.  This is a meaningful strengthening of the
adversary: indeed, it seems imprudent to assume that the adversary's
knowledge of the channel is limited to whatever is obtainable by black-box
queries (for instance, the adversary has some idea of a reasonable email
message or photograph should look like).  It does not
contradict our focus on black-box steganography: it is prudent for the
honest parties to avoid relying on particular properties of the channel,
while it is perfectly sensible for the adversary, in trying to break the
stegosystem, to take advantage of whatever information about the channel is
available.

\subsection{Pseudorandom Functions}
We use pseudorandom functions~\cite{GGM86} as a tool.  Because the
adversary in our setting has access to the channel, any cryptographic
tool used must be secure even given the information provided by the
channel.  Thus, the underlying assumption for our constructions is the existence of
pseudorandom functions that are secure given the channel oracle, which
is equivalent~\cite{HILL99} to the existence of one-way functions that
are secure given the channel oracle.  This is the minimal assumption
needed for steganography~\cite{HLvA02cor}.

Let $\F=\{F_\seed\}_{\seed\in\{0,1\}^*}$ 
be a family of functions, all with the same domain and range.
For a probabilistic adversary $A$, and channel $\CC$
with sampling oracle $M$, the \emph{PRF-advantage of
  $A$ over $\F$} is defined as
\[
\Adv^\mathrm{PRF}_{\F(n),\CC}(A)=
\left|\Pr_{\seed \gets \{0,1\}^n}[A^{M,F_\seed (\cdot)}=1]-
   \Pr_{g}[A^{M,g(\cdot)}=1]\right|\,,
\]
where $g$ is a random function with the same domain and range.
For a given $(t, d, q)$, the {\em insecurity} of a pseudorandom
function
family $\F$ with
respect to channel $\CC$ is defined as
\[\InSec^\mathrm{PRF}_{\F(n),\CC}(t,d,q)=
   \max_{(t,d,q) \mbox{ adversary } A} \{{\bf
     Adv}^\mathrm{SS}_{\F(n),\CC}(A)\}\,,
\]
where the maximum is taken over all adversaries that run in expected time
$t$, whose description size is at most $d$, and that make an expected
number of $q$ queries to their oracles.

The existence of pseudorandom functions is also the underlying
assumption for our lower bound; however, for the lower bound, we do
not need to give the adversary access to a channel oracle (because we
construct the channel).  To
distinguish this weaker assumption, we will omit the subscript $\CC$
from $\InSec$.

\section{The Lower Bound}

Recall that we define the rate of a stegosystem as the \emph{average number
of hiddentext bits per document sent} (this should not be confused with the
average number of hiddentext bits per \emph{bit} sent; note also that this
is the sender's rate, not the rate of information actually decoded by the
receiver, which is lower due to unreliability).  We set
out to prove that a reliable stegosystem with black-box access to the
channel with rate $w$ must make roughly $l2^w$ queries to the channel to
send a message of length $lw$.  Intuitively, this should be true because
each document carries $w$ bits of information on average, but since the
encoder knows nothing about the channel, it must keep on sampling until it
gets the encoding of those $w$ bits, which amounts to $2^w$ samples on
average.

In particular, for the purposes of this lower bound it suffices to
consider a restricted class of channels: the distribution of the
sample depends only on the length of the history (not on its
contents).  We will write $D_1, D_2, ..., D_i,...$, instead of
$D_\his$, where $i$ is the length of the history $\his$.
Furthermore, it will suffice for us to consider only distributions
$D_i$ that are uniform on a subset of $\Sigma$.  We will use
the notation $D_i$ both for the distribution 
and for the subset (as is often done for
uniform distributions).

Let $H$ denote the number of elements of $D_i$ (note that $H=|D_i|=2^h$), and let $S=|\Sigma|$.
Because the encoder  knows
the min-entropy $h$ of the channel, if $H=S$, then the encoder
knows the channel completely (it is simply uniform on $\Sigma$).  Therefore,
if $H=S$, then
there is no meaningful lower bound on the number of queries made
by the encoder to the channel oracle, because it does not need
to make any queries in order to sample from the channel.
Thus, we require that $H<S$ (our bounds will depend slightly
on the ratio of $S$ to $S-H$).

Our proof proceeds in two parts.  First, we consider a stegoencoder
$\SE$ that does not output anything that it did not receive as a
response from the channel-sampling oracle (intuitively, every good
stegoencoder should work this way, because otherwise it may output
something that is not in the channel, and thus be detected).  To be reliable---that is, to
find a set of documents that decode to the desired message---such an
encoder has to make many queries, as shown in Lemma~\ref{lemma-tree}.
Second, we formalize the intuition that a good stegoencoder should
output only documents it received from the channel-sampling oracle: we
show  that to be secure (i.e., not output something easily
detectable by the adversary), a black-box
$\SE$ cannot output anything it did not receive from the
oracle: if it does, it has an $1-H/S$ chance of being detected.

The second half of the proof is somewhat complicated by the fact that
we want to assume security only against bounded adversaries: namely,
ones whose description size and running time are polynomial in the
description size and running time of the encoder (in particular,
polynomial in $\log S$ rather than $S$).  
Thus, the adversary cannot be detecting a bad stegoenconder by simply
carrying a list of all the entries in $D_i$ for each $i$
and checking if the $i$th
document sent by the stegoencoder is in $D_i$, because that would make
the adversary's description too long.

This requires us to come up
with pseudorandom subsets $D_i$ of $\Sigma$ that have concise descriptions
and high min-entropy and whose membership is impossible for the
stegoencoder to predict.  In order to do that, we utilize techniques from
the truthful implementation of a boolean function with interval-sum
queries of \cite{GGN03} (truthfulness is important, because min-entropy
has to be high unconditionally).

\subsection{Lower Bound When Only Query Results Are Output}
If $D_1, D_2, \dots$ are subsets of
$\Sigma$, then we write $\vec{D} = D_1 \times D_2 \times \dots$ to denote
the channel that, on history length $i$, outputs a uniformly random
element of $D_i$.  If $|D_1| = |D_2| = \dots = 2^h$ then we say that
$\vec{D}$ is a {\em flat $h$-channel}.  We will consider flat $h$-channels.

Normally, one would think of
the channel sampling oracle for $\vec{D}$ as making a fresh random
choice from $D_i$ when queried on history length $i$.  However,
from the point of view of the stegoencoder, it does not matter
if the choice was made by the oracle in response to the query,
or before the query was even made.  It will be easier
for us to think 
of the oracle as having 
already made and written down countably many samples from
each $D_i$.  We will denote the $j$th sample from $D_i$ by $s_{i,j}$.
Thus, suppose that the oracle has already chosen
\begin{tabbing}
suppose\=\kill
\>$s_{1, 1}$, $s_{1, 2}, \dots, s_{1,
  j}, \dots$ \= from \= $D_1$, \\
\> $s_{2, 1}$, $s_{2, 2}, \dots, s_{2,
  j}, \dots$ \> from\> $D_2$, \\
\> \> \dots,\\
\> $s_{i, 1}$, $s_{i, 2}, \dots, s_{i,
  j}, \dots$ \>from\> $D_i$,\\
\>\> \dots \,.
\end{tabbing}
\noindent
We will denote the string containing all these samples by $\qs$ and
refer to it as a {\em draw sequence} from the channel.  We will give
our stegoencoder access to an oracle (also denoted by $\qs$) that, each
time it is queried with $i$, returns the next symbol from the sequence
$s_{i,1}, s_{i,2}, \dots, s_{i,j}, \dots$.  Choosing $\qs$
at random and giving the stegoencoder access to it is equivalent to
giving the encoder access to the usual channel-sampling oracle $M$ for
our channel $\vec{D}$.

Denote the stegoencoder's output by
$\SE^\qs(K,m,\his)=t=t_1t_2\dots t_l$, where $t_i\in \Sigma$.
Because we assume in this section that the stegoencoder outputs only
documents it got from the channel oracle, $t_i$ is an element of the
sequence $s_{i,1}, s_{i,2}, \dots, s_{i,j}, \dots$.   If $t_i$ is the
$j$th element of this sequence, then it took $j$ queries to produce
it.  We will denote by {\em weight of $t$ with respect to $\qs$}
the number of queries it took to produce $t$: $W(t, \qs) = 
\sum_{i=1}^l \min\{j  \suchthat s_{i, j} = t_i\}$.
In the next lemma, we
prove (by looking at the \emph{decoder}) that for any $\qs$ most
messages have high weight, i.e., must take many queries to encode.

\begin{lemma}
  \label{lemma-tree}
  Let $F : \Sigma^* \rightarrow \bool^*$ be an arbitrary (possibly
  unbounded) deterministic stegodecoder that takes a sequence
  $t\in\Sigma^l$ and outputs a message $m$ of length $lw$ bits.

  Then the probability that a random $lw$-bit message has an encoding of
  weight significantly less than $(1/e)l2^w$ is small.
  More precisely, for any
  $\qs \in \Sigma^{**}$ and any $N\in \nat$:
\[ 
  \pr_{m \in \bool^{lw}} [(\exists t \in \Sigma^l)(F(t) = m 
  \aand W(t,\qs)\leq N)] \le \frac{{N\choose l}}{2^{lw}} < 
  \left(\frac{Ne}{l2^w}\right)^l\,.
\]
\end{lemma}

\begin{proof}
Simple combinatorics show that the number of different sequences $t$
that have weight at most $N$ (and hence the number of messages
that have encodings of weight at most $N$) 
is at most ${N \choose l}$: indeed, it is simply the
number of positive integer solutions to 
$j_1  + \dots + j_l \le N$, which is the number of ways to put $l$
bars among $N-l$ stars (the number of stars to the right of the $i$th
bar corresponds to $j_i-1$), or, equivalently, the number of ways choose
$l$ positions out of $N$.  The total number of messages is $2^{lw}$.
The last inequality follows from ${N\choose l} <
\left(\frac{Ne}{l}\right)^l$ (which is a standard combinatorics fact
and follows from $k!\ge (k/e)^k$, which in turn follows by induction
on $k$ from $e>(1+1/k)^k$).
\end{proof}

Our lower bound applies when a stegosystem is used to encode messages
drawn uniformly from bit strings of equal length.  It can easily be
extended to messages drawn from a uniform distribution on any set.  If
the messages are not drawn from a uniform distribution, then, in
principle, they can be compressed before transmission, thus requiring
less work on the part of the stegoencoder.  We do not provide a lower
bound in such a case, because any such lower bound would depend on the
compressibility of the message source.

\subsection{Secure Stegosystems Almost Always Output Query Answers}
\label{section-multiplesources}
The next step is to prove that the encoder of a secure black-box
stegosystem must output only what it gets from the oracle, or else
it has a high probability of outputting something not in the channel.
Assume that $\vec{D}$ is a flat $h$-channel chosen uniformly at
random. 
For $t=t_1\dots t_l \in \Sigma^*$,
let $t\in \vec{D}$ denote that $t_i$ is in $D_i$
for each $i$.
In the following lemma, we demonstrate that, if the encoder's output
$t$ contains a document that it did not receive as a response
to a query, the chances that $t\in \vec{D}$ are at most
$H/S$. 

Before stating the lemma, we define the set $E$ of all possible flat
$h$-channels and draw sequences consistent with them: $E = \{
(\vec{D}, \qs) \suchthat s_{i,j} \in D_i\}$.  We will be taking
probabilities over $E$.  Strictly speaking, $E$
is an infinite set, because we defined $\vec{D}$ to be countable and
$\qs$ to have countably many samples from each $D_i$.  For clarity, it
may be
easiest
to think of truncating these countable sequences to a sufficiently
large value beyond which no stegoencoder will ever go, thus making $E$
finite, and then use
the uniform distribution on $E$.  Formally, $E$ can be defined as a
product of countably many discrete probability spaces (see,
e.g.,~\cite[Section 9.6]{FG97}), with uniform distribution on each.

\begin{lemma}
  \label{lemma-multiplesources}
  Consider any deterministic
  procedure $A$ that is given oracle access to a random flat
  $h$-channel $\vec{D}$ and outputs $t=t_1 t_2 \dots t_l \in \Sigma^*$
  (think of $A$ as the stegoencoder running on some input
   key, message, channel history, and fixed randomness).
  Provided that $h$ is sufficiently smaller than
  $\log S$, if $A$ outputs something it did not get from the oracle,
  then the probability $t\in \vec{D}$ is low.

  More precisely, let $Q_i$ be the set of responses $A$ received
  to its queries from the $i$th channel $D_i$.
  Define the following two events:
  \begin{itemize}
  \item {\underline n}on{\underline q}ueried:
    $\nq = \{ (\vec{D}, \qs) \in E\suchthat (\exists i) t_i \notin
    Q_i \}$
  \item \underline{in} \underline{s}upport:
    $\ins = \{ (\vec{D}, \qs) \in E \suchthat t \in \vec{D}\}$
  \end{itemize}
 Then: 
  \[ \pr_{(\vec{D}, \qs) \in E} [\ins \aand \nq]
     \leq \frac{H}{S} \,.\]
\end{lemma}
\begin{proof}
  If $A$ were always outputting just a single value ($l=1$), the proof would
  be trivial: seeing some samples from a random $D_1$
  does not help $A$ come up with another value from $D_1$, and $D_1$
  makes up only an $H/S$ fraction of all possible outputs of $A$.  The proof
  below is  a generalization of this argument for $l\ge
  1$, with care to avoid simply taking the union bound, which would get us
  $lH/S$ instead of $H/S$.

  Let $\nq_i = \{(\vec{D}, \qs)\in E \suchthat t_1\in Q_1, t_2\in Q_2 ,
  \dots, t_{i-1}\in Q_{i-1}, t_i \notin Q_i\}$
  be the event $t_i$ is the first element of the output that
  was
  not returned by
  the oracle as
  an answer to a query.
  Observe that $\bigcup_i \nq_i = \nq$ and that $\nq_i$ are
  disjoint events and, therefore, $\sum_i \Pr[\nq_i]=1$.
  Now the probability we are interested in is
\[
\Pr[\ins\aand \nq] = \sum_i \Pr[\ins \aand \nq_i] =
  \sum_i \Pr[\ins \suchthat \nq_i]\Pr[\nq_i]\,.
\]
To bound $\Pr[\ins \suchthat \nq_i]$, fix any
\begin{tabbing}
suppose\=\kill
\>$\qs=\ $\= $s_{1, 1}$, $s_{1, 2},$\=$\dots, s_{1,
  q_1}$\=, \\
\>\> $s_{2, 1}$, $s_{2, 2},$\>$\dots, s_{2,
  q_2}$\>, \\
\>\>\> \dots\> ,
\end{tabbing}
\noindent
such
that $A^\qs$ asks exactly $q_1$ queries from $D_1$, $q_2$ queries
from $D_2$, \dots.  Note that such $\qs$ determines the behavior
of $A$, including its output.  Assume that, for this $\qs$, the
event $\nq_i$ happens.  We will take the probability $\Pr[\ins \suchthat \nq_i]$
over a random $\vec{D}$ consistent with $\qs$ (i.e., for which
$s_{1,1}, s_{1,2}, \dots s_{1, q_1}\in D_1, s_{2,1}, s_{2,2}\dots
s_{2, q_2}\in D_2, \dots$).   This probability can be computed simply
as follows: if $q'_i$ is the number of distinct elements in $s_{i,1}, s_{i,2},
\dots, s_{i, q_i}$, then  there are $ S-q'_i \choose H-q'_i$ equally
likely
choices for $D_i$
(because $q'_i$ elements of $D_i$ are already determined).  However,
for $\ins$ to happen, $D_i$ must also contain $t_i$,
which is not among $s_{i,1}, s_{i,2},
\dots, s_{i, q_i}$ (because we assumed $\nq_i$ happens).  The choices
of $D_1, \dots, D_{i-1}, D_{i+1}, \dots$ do not matter. Therefore,
\[
\Pr[\ins \suchthat \nq_i]=
 \frac{{S-q'_i-1 \choose H-q'_i-1}}{{S-q'_i \choose H-q'_i}} =
  \frac{H-q'_i}{S-q'_i} \leq \frac{H}{S}\,.
\]

The above probability is for \emph{any} fixed $\qs$ of the right
length
and randomly chosen
$\vec{D}$ consistent with $\qs$.  Therefore, it also holds for
randomly chosen $(\vec{D}, \qs)\in E$, because the order in which
$\qs$ and $\vec{D}$ are chosen and the values in $\qs$ beyond
what
$A$ queries do not affect the probability.  We thus have
\[
\Pr_{(\vec{D}, \qs)\in E}[\ins\aand \nq] =
\sum_i \Pr[\ins \suchthat \nq_i]\Pr[\nq_i] \leq
\sum_i \frac{H}{S} \Pr[\nq_i] = \frac{H}{S}\,.
\]
\end{proof}

\subsection{Lower Bound for Unbounded Adversary}

We now want to tie together Lemmas \ref{lemma-tree} and
\ref{lemma-multiplesources} to come up with a lower bound on the
efficiency of the stegoencoder in terms of rate, reliability, and
security. Note that some work is needed, because even though
Lemma~\ref{lemma-tree} is about reliability and
Lemma~\ref{lemma-multiplesources} is about security, neither mentions
the parameters $\Rel$ and $\InSec$.

Assume, for now,  that the adversary can test whether $t_i$ is in the support of
$D_i$.  (This is not possible if $D_i$ is completely random and the adversary's
description is small compared to $S=|\Sigma|$; however, it serves as a
useful warm-up for the next section.)
Then, using Lemma~\ref{lemma-multiplesources}, 
it is easily shown that, if the stegoencoder has insecurity
$\epsilon$, then it cannot output something it did not receive as response
to a query with probability higher than $\epsilon /(1-H/S)$.   This
leads to  the following theorem.

\begin{theorem}
  \label{theorem-main-unbounded}
  Let $(\SE,\SD)$ be a black-box stegosystem with insecurity $\epsilon$
  against an adversary who has an oracle for testing membership in the
  support of $\CC$,
  unreliability $\rho$ and rate $w$ for an alphabet $\Sigma$ of size $S$.
  Then, for any positive integer $H<S$, there exists a channel with min-entropy $h=\log_2 H$ such that
  the probability
  that the encoder makes at most $N$ queries to send a random message of length $lw$
  is at most
  \[
     \left(\frac{Ne}{l2^w}\right)^l + \rho + \epsilon R\,,
  \] 
  and the expected number of queries per stegotext symbol is therefore at least 
  \[
    \frac{2^w}{e}\left(\frac{1}{2}- \rho - \epsilon R\right)\,,
  \]
  where $R=S/(S-H)$.
\end{theorem}

Note that, like Lemma~\ref{lemma-tree}, this theorem and
Theorem~\ref{theorem-main-bounded} apply when a stegosystem is used to
encode messages drawn uniformly from the distribution of all $lw$-bit
messages (see remark following the proof of Lemma~\ref{lemma-tree}).

\begin{proof}

We define the following events, which are all subsets of $E \times
   \bool^* \times \bool^{lw}\times \bool^*$ (below $v$ denotes the randomness of $\SE$):
  \begin{itemize}
  \item ``$\SE$ makes \underline{few}  queries to encode $m$ under $\key$'':
    $\few = \{\vec{D}, \qs, \key, m, v\suchthat \SE^\qs(\key, m; v)
\allowbreak \mbox{ makes
\allowbreak at \allowbreak most \allowbreak $N$ \allowbreak
    queries}\}$ (note that this is the event whose probability we want
  to bound)
  \item ``$\SE$ outputs a \underline{corr}ect encoding of $m$ under $\key$'':
    $\corr = \{\vec{D}, \qs, \key, m, v\suchthat \SD(\key, \SE^\qs(\key, m;
  v)) \allowbreak = m \}$
  \item ``$m$ has an encoding $t$ under $\key$, and this encoding has
    \underline{low} weight'':
    $\low = \{ \vec{D}, \qs, \key, m, v(\exists t)\suchthat \allowbreak  \SD(\key, t) = m \aand W(t,\qs) \leq N \}$
  \item $\ins$ and $\nq$ as in Lemma \ref{lemma-multiplesources}, but as
    subsets of $E \times \bool^* \times \bool^{lw}\times \bool^*$
  \end{itemize}
  Suppose that $\SE$ outputs a correct encoding of a message $m$. In
  that case, the probability that it made at most $N$ queries to the
  channel is upper bounded by the probability that: (i) there exists an
  encoding of $m$ of weight at most $N$, or (ii) $\SE$ output something it
  did not query. In other words,
  \[ \pr [\few \suchthat \corr] \leq
  \pr [\low \suchthat \corr] + \pr [\nq \suchthat \corr] . \]
  Now we have
  \begin{eqnarray*}
    \pr[\few] & = & \pr[\few \cap \corr] + \pr[\few \cap \overline{\corr}] \\
    & \leq & \pr[\few \cap \corr] + \pr[\overline{\corr}] \\
    & = & \pr[\few \suchthat \corr] \cdot \pr[\corr] + \pr[\overline{\corr}] \\
    & \leq & (\pr[\low \suchthat \corr] + \pr[\nq \suchthat \corr]) \cdot
    \pr[\corr] + \pr[\overline{\corr}] \\
    & = & \pr[\low \cap \corr] + \pr[\nq \cap \corr] + \pr[\overline{\corr}] \\
    & \leq & \pr[\low] + \pr[\nq] + \pr[\overline{\corr}]\,.
  \end{eqnarray*}
  
  Because insecurity is $\epsilon$, $\Pr[\overline{\ins}]\le
  \epsilon$.  Hence,
  \begin{equation}
    \label{equation-few2}
    \Pr[\nq] = \frac{\Pr[\overline{\ins}\cap\nq]}{\Pr[\overline{\ins}\suchthat{\nq}]}= \frac{\Pr[\overline{\ins}]}{\Pr[\overline{\ins}\suchthat
  \nq]}\leq \frac{\epsilon}{1-H/S} \, 
  \end{equation}
  (the second equality follows from the fact that 
   if the encoder outputs something not in $\vec{D}$, then
  it must have not queried it, i.e., $\overline{\ins} \subseteq \nq$; 
  the inequality follows from Lemma~\ref{lemma-multiplesources}).

  By Lemma \ref{lemma-tree} we have
  \begin{equation}
    \label{equation-few3}
    \pr[\low] \leq \left(\frac{Ne}{l2^w}\right)^l\, .
  \end{equation}

  Now by combining (\ref{equation-few2}),
  (\ref{equation-few3}), and the fact that $\Pr[\overline{\corr}]\le
  \rho$ by reliability,
  we get that
  \[ \pr [\few] \leq \left(\frac{Ne}{l2^w}\right)^l  + \rho + \frac{\epsilon}{1-H/S}\, . \]

Note that the probability is taken, in particular, over a random choice of
$\vec{D}$.  Therefore, it holds for at least one flat $h$-channel.

Let random variable $q$ be equal to the number of queries made by $\SE$ to
encode $m$ under $K$. 
Then, letting $d=l2^w/e$ and $c=1-\rho-\frac{\epsilon}{1-H/S}$, we get
\begin{eqnarray*}
 \expc{q} & = & \sum_{N \geq 0} \pr[q > N]
   \ge  \sum_{N=0}^{\ceil{d}-1}
                    c - \left( \frac{N}{d} \right)^l 
\ge  \sum_{N=0}^{\ceil{d}-1}
                    c - \frac{N}{d} 
 =  c\ceil{d} - \frac{(\ceil{d}-1)\ceil{d}}{2d} \geq  \left(c-\frac{1}{2}\right)\ceil{d}\,.
\end{eqnarray*}
The expected number of queries per document sent is $(\expc{q}) / l$
and so is at least
$(\frac{1}{2}-\rho-\frac{\epsilon}{1-H/S})(2^w/e)$.
\end{proof}

\subsection{Lower Bound for Computationally Bounded Parties}
\label{section-bounded-adversaries}
We now want to establish the same lower bound without making such a strong
assumption about the security of the stegosystem.  Namely, we do not want
to assume that the insecurity $\epsilon$ is low unless the adversary's
description size and running time are feasible (``feasible,'' when made rigorous,
will mean some fixed polynomial in the description size and running time
of the stegoencode and in a security parameter for a function
that is pseudorandom against the stegoencoder).  Recall that our
definitions allow the adversary to depend on the channel; thus, our goal is
to construct channels that have short descriptions for the adversary but
look like random flat $h$-channels to the black-box stegoencoder.  In other
words, we wish to replace a random flat $h$-channel with a pseudorandom
one.

We note that the channel is pseudorandom only in the sense that it has a
short description, so as to allow the adversary to be computationally
bounded.  The min-entropy guarantee, however, can not be replaced with a
``pseudo-guarantee'': else the encoder is being lied to, and our lower
bound is no longer meaningful.  Thus, a simpleminded approach, such as
using a pseudorandom predicate with bias $H/S$ applied to each symbol and
history length to determine whether the symbol is in the support of the
channel, will not work here: because $S$ is constant, eventually (for some
history length) the channel will have lower than guaranteed min-entropy
(moreover, we do not wish to assume that $S$ is large in order to
demonstrate that this is unlikely to happen; our lower bound should work
for any alphabet).
Rather, we need the pseudorandom implementation of the channel to be
truthful\footnote{In this case, truthfulness implies that for each history
length, the support of the channel has exactly $H$ elements.} in the sense
of \cite{GGN03}, and so rely on the techniques developed therein.

The result is the following theorem, which is similar to
Theorem~\ref{theorem-main-unbounded}, except for a small term
introduced by pseudorandomness of the channel.

\begin{theorem}
  \label{theorem-main-bounded} 
  There exist polynomials $p_1, p_2$ and constants $c_1, c_2$ with the following
  property.
  Let $\st(\kappa)$ be a black-box stegosystem with security parameter
  $\kappa$, description size $\delta$,
  unreliability $\rho$, rate $w$, and running time $\tau$ 
  for the alphabet $\Sigma=\{0,1,\dots, S-1\}$.
  Assume that there
  exists a  pseudorandom function family $\F(n)$ with insecurity
  $\InSec^\mathrm{PRF}_{\F(n)}(t,d,q)$.  Then, for any positive integer $H<S$,
  there exists a  channel $\CC$  with min-entropy $h=\log_2 H$ such that the probability
  that the encoder makes at most $N$ queries to send a random message of length $lw$
  is upper bounded by
  \[
     \left(\frac{Ne}{l2^w}\right)^l + \rho + 
     R\epsilon + 
     (R+1)\left(\InSec^\mathrm{PRF}_{\F(n)} (p_1(\tau,n), \delta+c_1,
  p_1(\tau,n))+\tau 2^{-n}\right)\,,
  \] 
  and the expected number of queries per stegotext symbol is therefore at least 
  \[
    \frac{2^w}{e}\left(\frac{1}{2}-\rho - 
     R\epsilon - 
     (R+1)\left(\InSec^\mathrm{PRF}_{\F(n)} (p_1(\tau,n), \delta+c_1,
  p_1(\tau,n))+\tau 2^{-n}\right)\right)\,,
  \]
  where $R=S/(S-H)$ and $\epsilon$ is the
  insecurity the stegosystem $\st$ on the channel $\CC$ against
  adversaries running in time $p_2(n,\log S, n)$ of description size $n+c_2$,
  making just one query of length $lw$ to $\SE$ or $O$ (i.e.,  $\epsilon=
  \InSec^\mathrm{SS}_{\st(\kappa),\CC} (p_2(n, \log S, l),n+c_2,1,lw)$).
\end{theorem}

\begin{proof}
The main challenge lies in formulating the analogue of Lemma
\ref{lemma-multiplesources} under computational restrictions.  Lemma
\ref{lemma-multiplesources} and its use in
Theorem~\ref{theorem-main-unbounded}
relied on: (i) the inability of the
encoder to predict the behavior of the channel (because the channel is
random) and (ii) the ability of the adversary to test if a given string is
in the support of the channel (which the adversary has because it is
unbounded).  We need to mimic this in the computationally bounded case.  We
do so by constructing a channel whose support (i) appears random to a
bounded encoder, but (ii) has an efficient test of membership that the
adversary can perform given only a short advice. As already mentioned, we
wish to replace a random channel with a pseudorandom one and give the short
pseudorandom seed to the adversary, while keeping the
min-entropy guarantee truthful.

The next few paragraphs will explain how this is done, using the
techniques of huge random objects from~\cite{GGN03}.  A reader not
familiar with~\cite{GGN03} may find it easier to skip to the paragraph
entitled ``Properties of the Pseudorandom Flat-$h$ Channels,'' where
the results of this---i.e., the properties of the channel that we
obtain---are summarized.

\paragraph{Specifying and Implementing the Flat-$h$ Channel}
For the next few paragraphs, familiarity with~\cite{GGN03} will be assumed.
Recall that~\cite{GGN03} requires a specification of the object that
will be pseudorandomly implemented, in the form of a Turing machine
with a countably infinite random tape. 
It would be straightforward to specify the
channel as a random object (random subset $D$ of $\Sigma$ of size $H$)
admitting two types of queries: ``sample'' and ``test membership.''  But
a pseudorandom implementation of such an
object would also replace random sampling with pseudorandom sampling,
whereas in a stegosystem the encoder is guaranteed a truly random sample
from $D$ (indeed, without such a guarantee, the min-entropy guarantee is no
longer meaningful).  Therefore, we need to construct a slightly different
random object, implement it pseudorandomly, and add random sampling on top
of it.  We specify the random object as follows.  Recall that $S=|\Sigma|$,
$h$ is the min-entropy, and $H=2^h$.

\begin{definition}[Specification of a flat $h$-channel]
  \label{definition-specification-flat}
  Let $M_{\omega}$ be a probabilistic Turing machine with
  an infinite random tape $\omega$. On
  input five integers $(S, H, i, a, b)$, (where $0<H\le S$, $i>0$,
  $0\le a\le b<S$), $M_{\omega}$ does the following:
  \begin{itemize}
  \item divides $\omega$ into consecutive substrings $y_1, y_2, \dots$ 
    of length $S$ each;
  \item identifies among them the substrings that have exactly $H$ ones;
    let $y$ be the $i$th such substring (with probability one, there are
    infinitely many such substrings, of course);
  \item returns the number of ones in $y$ between, and including, positions
    $a$ and $b$ in $y$ (positions are counted from $0$ to $S-1$).
  \end{itemize}
\end{definition}

In what way does $M = M_{\omega}$ specify a flat
$h$-channel? To see that, identify $\Sigma$ with $\{0, \dots, S-1\}$,
and let $D_i$ be the subset of $\Sigma$ indicated by the ones in $y$.
Then $D_i$ has cardinality $H$ and testing membership in $D_i$
can be realized using a single
query to $M$:
\begin{tt}
  \begin{tabbing}
    123\=123\=\kill
    $\insupp^M$($i$, $s$): \\
    \> return $M(S, H, i, s, s)$
  \end{tabbing}
\end{tt}
Obviously, $D_i$ are selected
uniformly at random and independently of each other.  
Thus, this object specifies the correct channel and allows membership
testing.

We now use this object to allow for random sampling of $D_i$.
Outputting a random element of $D_i$ can be realized via $\log S$ queries
to $M$, using the following procedure (essentially, binary search):
\begin{tt}
  \begin{tabbing}
    123\=123\=123\=\kill
    $\rndelt^M$($i$): \\
    \> return random-element-in-range$^M$($S, H, i, 0, S-1$) \\
    \\
    random-element-in-range$^M$($S, H, i, a, b$): \\
    \> if $a = b$ then return $a$ and terminate \\
    \> $\mathit{mid} \leftarrow \lfloor (a+b)/2 \rfloor$ \\
    \> $\mathit{total} \leftarrow M(S, H, i, a, b)$ \\
    \> $\mathit{left} \leftarrow M(S, H, i, a, \mathit{mid})$ \\
    \> $r \leftrand \{1, \dots, \mathit{total}\}$ \\
    \> if $r \leq \mathit{left}$ then \\
    \>\> random-element-in-range$^M$($S, H, i, a, \mathit{mid}$) \\
    \> else \\
    \>\> random-element-in-range$^M$($S, H, i, \mathit{mid}+1, b$)
  \end{tabbing}
\end{tt}

We can implement this random object pseudorandomly using the same
techniques as~\cite{GGN03} uses for implementing random boolean
functions with interval sums (see \cite[Theorem 3.2]{GGN03}).  Namely,
the authors of \cite{GGN03} give a construction of a truthful
  pseudo-implementation of a random object determined by a random boolean
  function $f : \{0, \dots, 2^n-1\} \rightarrow \bool$ that accepts
    queries in the form of two $n$-bit integers
  $(a, b)$ and answers with $\sum_{j=a}^b f(j)$. Roughly, their
  construction is as follows. Let $S=2^n$.
  Imagine a full binary tree of depth $n$,
  whose leaves contain values $f(0), f(1), \dots, f(S-1)$. Any other
  node in the tree contains the sum of leaves in its subtree. Given
  access to such tree, we can compute any sum $f(a) + f(a+1) + \dots
  + f(b)$ in time proportional to $n$.
  Moreover, such trees need not be stored fully but can be
  evaluated dynamically, from the root down to the leaves, as follows.
  The value in the root (i.e., the sum of all leaves)
  has binomial distribution and can be filled in pseudorandomly.  Other
  nodes have more complex distributions but can be also filled in
  pseudorandomly and consistently, so that they contain the sums of
  their leaves.  The construction uses a pseudorandom function to come
  up with the value at each node.

  We need to make three modifications.   First,
  we simply fix
  the value in the root to $H$, so that $f(0) + f(1) + \dots + f(S-1) =
    H$.  Second, we allow $S$ to be not a power of 2.  Third, in
  order to create multiple distributions $D_i$, we add $i$ as
  an input to the pseudorandom
  function, thus getting different (and
  independent-looking) randomness for each $D_i$.

  Having made these modifications, we obtain a truthful pseudo-implementation
  of $M$.
  It can be used within
  $\insupp$ and $\rndelt$ instead of $M$, for
  efficient membership testing and truly random sampling from our
  pseudorandom channel.  

\paragraph{Properties of the Pseudorandom Flat $h$-Channels}
We thus obtain that, given a short random seed $\omega$, it is
possible to create a flat $h$-channel that is indistinguishable from
random and allows for efficient membership testing
and truly
random sampling given $\omega$.  
To emphasize the pseudorandomness of the channel, in our notation
we will use $\dpr$
insted of $D$ and keep the seed $\omega$ explicit as a supercript.  Thus,
$\dpr_i^\omega$ is a pseudorandom subset of $\Sigma$ of size $H$,
and the channel is denoted by $\overrightarrow{\dpr}^\omega =
\dpr^\omega_1 \times \dpr^\omega_2 \times \dots$.  
Similarly to $E$
defined in
Section~\ref{section-multiplesources} for truly random channels, define
$\epr_n = \{ (\omega, \qs) \suchthat |\omega|=n, s_{i,j} \in \dpr^\omega_i\}$.

Because $\overrightarrow{\dpr}^\omega$ has the requisite
min-entropy, it is valid to expect proper performance of the
stegoencoder on it; because it is pseudorandom, an analog of
Lemma~\ref{lemma-multiplesources} will still hold; and because it has
efficient membership testing given a short seed, the adversary will be
able to see if an output of the stegoencoder is not from it.

We are now ready to formally state the claim about the properties of
$\overrightarrow{DPR}$.  For this claim, and for the rest of the
proof,  we assume existence of a family of pseudorandom
functions $\F$ with insecurity $\InSec^\mathrm{PRF}_{\F(n)} (t, d, q)$
(recall that $\InSec$ is a bound on the distinguishing advantage of any
adversary running in time at most $t$ of description size at most $d$
making at most $q$ queries).   To simplify the notation, we will note that
for us $d$ always will be at most
description size of the stegosystem plus some constant $c_1$, and that $q\le t$.
We will then
write $\iprf(n, t)$ instead of $\InSec^\mathrm{PRF}_{\F(n)} (t, d,
q)$.

\begin{claim}
  \label{claim-implementation-hchannel-final}
  There is a polynomial $p$ and a family of channels
  $\overrightarrow{\dpr}^\omega$,
  indexed by a string $\omega$ of length $n$ (as well as values $H$
  and $S$), such that, for any
  positive integers
  $n, i$ and $H\le S$, channel $\overrightarrow{\dpr}$ has the
  following properties:
  \begin{itemize}
  \item is a flat $h$-channel for $h=\log H$ on the alphabet $\{0,
  \dots, S-1\}$;
  \item allows for sampling and membership testing in time
  polynomial in $n$, $\log S$, and $\log i$ given $\omega, i, H$, and $S$ as
  inputs;

\newcommand{\mem}{\mathit{Memb}}
 \item is pseudorandom in the following sense: for any $H$, $S$, and any
  oracle machine (distinguisher)
  $A$ with running time $\tau\ge \log S$,
  \[ 
    \left|\pr_{(\vec{D}, \qs) \leftarrow E} [A^{\qs, \mem(\vec{D})}() = 1] -
    \pr_{(\omega, \qs) \leftarrow \epr_n} [A^{\qs, \mem(\omega)}() = 1]\right| <
    \iprf(n, p(\tau,n))
    + \tau  2^{-n}\,,
  \]
  where $\mem(\vec{D})$ and $\mem(\omega)$ denote membership testing
  oracles for $\vec{D}$ and $\overrightarrow{\dpr}^\omega$, respectively.
  \end{itemize}
\end{claim}

The claim follows from the results of~\cite{GGN03}
with minor modifications, as presented above.  We present no proof here.

Note  that the second argument to $\iprf$ depends on $S$ only
to the extent $\tau$ does;
this is important, because, even for large alphabets and
high-entropy channels, we want to keep the second
argument to $\iprf$ as low a possible so that $\iprf$ is as low as
possible.

\paragraph{Stegosystems Running with $\dpr$ 
Almost Always Output Query Answers}
Having built pseudorandom channels, we now state the analog of
Lemma~\ref{lemma-multiplesources} that works for stegosystems secure
only against bounded adversaries.  Fix some $H$ and $S$.
Let $A$ be the same as in Lemma~\ref{lemma-multiplesources}, but given
access to $\overrightarrow{\dpr}^\omega$ instead of $\vec{D}$, and let
$t=t_1\dots t_l$ be its output and 
$Q_i$ be the set of responses $A$ received to its queries of
the $i$th channel $\dpr_i$.  Analogously to $\nq$ and $\ins$, 
define the following two families of events, indexed by $n$, the
security parameter for the PRF.
\begin{itemize}
\item {\underline n}on{\underline q}ueried,
\underline{p}seudo\underline{r}andom version: $\nqpr_{n} = \{(\omega, \qs) \in \epr_n 
  \suchthat (\exists i)t_i \notin Q_i \}$

\item \underline{in} \underline{s}upport,
\underline{p}seudo\underline{r}andom version: 
$\inspr_{n} = \{(\omega, \qs) \in \epr_n 
  \suchthat t \in \overrightarrow{\dpr}^\omega\}$
\end{itemize}

We show that high probability of $\inspr_{n}$ implies low probability of
$\nqpr_{n}$. Formal statement of the lemma follows.  To simplify the notation,
let $R=S/(S-H)$.
\begin{lemma}
  \label{lemma-nonqueried-computational}
  There exists a polynomial $p_1$ such that, for any $A$ running in time
  $\tau\ge \log S$,
  if $\pr[\overline{\inspr_{n}}] < \epsilon(n)$,
  then
  \[ \pr[\nqpr_{n}] < R\epsilon(n) + 
  (R+1)(\iprf(n, p_1(\tau,n))+\tau 2^{-n})\,. \]
\end{lemma}
\begin{proof}
Let $\ins$ and $\nq$ be the same as in
Lemma~\ref{lemma-multiplesources}.
Let $A'$ be a machine that is given an oracle which tests
membership in the channel.  Let $A'$ run $A$ to get $t$ and output $1$ if
and only if the membership oracle says that $t$ is in the channel. Applying
 Claim \ref{claim-implementation-hchannel-final} to $A'$,
  we have that for some polynomial $p'$ (namely, the polynomial
  $p(\tau+t_{A'}(\tau), n)$, where $t_{A'}$ is  the extra time that
  $A'$ needs to run after $A$ is finished),
  \[ |\pr[\inspr_{n}] - \pr[\ins]| < \iprf(n, p'(\tau, n)) + \tau 2^{-n} . \]
  Therefore $\pr[\overline{\ins}] < \epsilon(n) + \iprf(n,
  p(\tau+p'(\tau, n))) + \tau 2^{-n}$.
  It now follows, by the same derivation as for
  Equation~(\ref{equation-few2}) in the proof of Theorem
  \ref{theorem-main-unbounded},
  that
 \[
  \pr[\nq] < \frac{\epsilon(n) + \iprf(n, p'(\tau, n)) + \tau 2^{-n}}{1-H/S}\,.
 \]

  Let $A''$ be a machine that runs $A$ and outputs $1$ if and only if
  $A$ outputs something it did not receive as a query response.
  Applying Claim \ref{claim-implementation-hchannel-final} to $A''$, we get
  that, for some polynomial $p''$ (namely, the polynomial
  $p(\tau+t_{A''}(\tau), n)$, where $t_{A''}$ is  the extra time that
  $A''$ needs to run in addition to $A$), we get
  $|\pr[\nqpr_{n}] - \pr[\nq]| < \iprf(n, p''(\tau, n)) + \tau 2^{-n}$.
  Therefore,
  \[ \pr[\nqpr_{n}] < \frac{\epsilon(n) + \iprf(n, p'(\tau,n))}{1-H/S} +
  \iprf(n, p''(\tau, n)) + (1+R)\tau 2^{-n}\,.\]
  Now let $p_1\ge \max(p', p'')$.
\end{proof}

\paragraph{Completing the Proof.}
We are now ready to prove Theorem~\ref{theorem-main-bounded}.
We define the same events as in the proof of
Theorem~\ref{theorem-main-unbounded}, except as subsets of 
$\epr_n\times \bool^* \times \bool^{lw}\times \bool^*$ rather than
$E\times \bool^* \times \bool^{lw}\times \bool^*$ 
(we use the
suffix PR to emphasize that they are for the pseudorandom channel):
$\fewpr_n, \corrpr_n, \lowpr_n$ denote, respectively, that $\SE$ made at
most $N$ queries, that $\SD$ correctly decoded the hiddentext, and
that the hiddentext has a low-weight encoding.  

Just like in the proof of
\ref{theorem-main-unbounded}, it holds that  $\pr[\fewpr_{n}] \leq \pr[\lowpr_{n}] +
  \pr[\nqpr_{n}] + \pr[\overline{\corrpr_{n}}]$ and  that $\pr[\overline{\corrpr_{n}}] <
  \rho$ and $\pr[\lowpr_{n}] < (Ne/l2^w)^l$.
 It is left to argue a bound on $\pr[\nqpr_{n}]$.

  Consider an adversary against our stegosystem
  that contains $\omega$ as part of its description, gives its oracle
  a random message to encode, and then tests if the output
  is in $\overrightarrow{\dpr}^\omega$.  
  It can be implemented to run in
  $p_2(n, \log S, l)$ steps for some polynomial $p_2$
  and has description
  size $n+c_2$ for some constant $c_2$.
  Hence, its probability of detecting a
  stegoencoder output that is not in $\overrightarrow{\dpr}^\omega$ cannot
  be more than the insecurity
 $\epsilon=
  \InSec^\mathrm{SS}_{\st(\kappa),\overrightarrow{\dpr}^\omega}
  (p_2(n, \log S, l),n+c_2,1,lw)$. 
 In other words,
  $ \pr [\overline{\inspr_{n}}] \leq \epsilon$, and, by 
  Lemma \ref{lemma-nonqueried-computational}, we get
  \[ \pr [\nqpr_{n}] \leq R\epsilon+ (R+1)(\iprf(n,
  p_1(\tau,n))+\tau 2^{-n})\,. \]

  Finally, to compute a bound on the expected value, we apply the same
  method as in the proof of Theorem~\ref{theorem-main-unbounded}.
\end{proof}

\paragraph{Discussion.}
The proof of Theorem \ref{theorem-main-bounded} relies fundamentally on
Theorem \ref{theorem-main-unbounded}: specifically,
Lemma~\ref{lemma-nonqueried-computational} relies on Lemma~\ref{lemma-multiplesources}.  In other words, to prove a lower
bound in the computationally bounded setting, we use the corresponding
lower bound in the information-theoretic setting.
To do so, we replace an object of an exponentially large size (the channel)
with one that can be succinctly described. This replacement substitutes
{\em some} information-theoretic properties with their computational
counterparts.  However, for a lower bound to remain ``honest'' (i.e., not
restricted to uninteresting channels), some global properties must remain
information-theoretic.  This is where the truthfulness of huge random
objects of~\cite{GGN03} comes to the rescue.  We hope that other
interesting impossibility results can be proved in a similar fashion by
adapting an information-theoretic result using the paradigm
of~\cite{GGN03}.  We think truthfulness of the objects will be important in
such adaptations for the same reason it was important here.

Note that the gap in the capabilities of the adversary and
encoder/decoder is different in the two settings:
in the information-theoretic case, the adversary is given
unrestricted computational power, while in the computationally bounded
case, it is assumed to run in polynomial time but is given the secret
channel seed.  However, 
in the information-theoretic case, we may remove the
gap altogether by providing both the adversary and the encoder/decoder
with a channel membership oracle  and
still obtain a lower bound analogous\footnote{A lower bound on the number
of samples per document sent becomes trivially zero if the encoder is
given as much time as it pleases, in addition to the membership oracle of
the flat channel.  Yet it should not be difficult to prove that it must
then run for $O(2^w)$ steps per document sent.} to that of Theorem
\ref{theorem-main-bounded}. We see no such opportunity to remove the gap in the
computationally bounded case (e.g., equipping the encoder/decoder with
the channel seed seems to break our proof).  Removing this asymmetry in the
computationally bounded case  seems challenging and worth pursuing.

\section{The Stateful Construction $\stf$}
\label{section-stateful}
The construction $\stf$ relies on a pseudorandom function family
$\F$.  In addition to the security parameter $\kappa$ (the length
of the PRF key $K$), it depends on the rate parameter $w$.  Because it is
stateful, both encoder and decoder take a counter $\ctr$ as input.

Our encoder is similar to the rejection-sampler-based encoder
of~\cite{HLvA02cor} generalized to $w$ bits: it simply samples elements from
the channel until the pseudorandom function evaluated on the element
produces the $w$-bit symbol being encoded.  The crucial difference of our
construction is the following: to avoid introducing bias into the channel,
if the same element is sampled twice, the encoder simply flips a random coin to
decide whether to output that element with probability $2^{-w}$.  
Hopper~\cite[Construction 6.10]{Hop04} independently
proposes a similar construction, except instead of flipping a fresh random
coin, the encoder evaluates the pseudorandom function on a new counter
value (there is a separate counter associated to each sampled document,
indicating how many times the document has been sampled), thus conserving
randomness.

Observe that, assuming $\F$ is truly random rather than pseudorandom, each
sample from the channel has probability $2^{-w}$ of being output,
independent of anything else, because each time fresh randomness is being
used. Of course, this introduces unreliability,
which is related to the probability of drawing the same element from
$D_\his$ twice.

\medskip
\noindent
\begin{tabular}{ll}
\begin{minipage}{3.5in}
\begin{tabbing}
{\bf Procedure $\stf.\SE(K,w,m,\his, \ctr)$}:\\
123\=123\=123\=123\=123\=\kill
\>Let $m=m_1 m_2 \dots m_l$, where $|m_i|=w$
\hspace{1in}  \\
\>\kwfor $i\gets 1$ \kwto $l$:\\
\>\> $j\gets 0$; $\outflag\gets 0$; $\ctr\gets\ctr+1$\\
\>\> \kwrepeat:\\
\>\>\>$j\gets j+1$\\
\>\>\>$s_{i,j} \gets M(\his)$\\
\>\>\>\kwif $\exists j'<j$ s.t. $s_{i,j}=s_{i,j'}$\\
\>\>\>\> \kwlet $c\in_R \{0,1\}^w$ \\
\>\>\>\> \kwif $c=m_i$ \kwthen $\outflag \gets 1$\\
\>\>\>\kwelse \kwif $F_K(\ctr,s_{i,j})=m_i$ \\
\>\>\>\> \kwthen $\outflag\gets 1$\\
\>\> \kwuntil $\outflag=1$\\
\>\> $s_i\gets s_{i,j}$; $\his\gets \his || s_i$\\
\> \kwoutput $s=s_1 s_2 \dots s_l$\\
\end{tabbing}
\end{minipage}
&
\begin{minipage}{3in}
\begin{tabbing}
123\=123\=123\=\kill
{\bf Procedure $\stf.\SD(K,w,s,\ctr)$}:\\
\> Let $s=s_1\dots s_l$, where $s_i\in\Sigma$\\
\>\kwfor $i=1$ \kwto $l$\\
\>\>$\ctr\gets\ctr+1$\\
\>\>$m_i \gets F_K(\ctr,s_i)$\\
\> \kwoutput $m = m_1m_2\cdots m_l$\\
\\
\\
\\
\\
\\
\\
\\
\\
\\
\end{tabbing}
\end{minipage}
\end{tabular}

\begin{theorem}
The stegosystem $\stf$ has insecurity 
${\bf InSec}^\mathrm{SS}_{\stf(\kappa,w)} (t, d, l, lw)=
{\bf InSec}^\mathrm{PRF}_{\F(\kappa)} (t+O(1),\allowbreak d+O(1),\allowbreak l2^w)$.  For each $i$,
the probability that $s_i$ is decoded incorrectly is $2^{-h+w}+
{\bf InSec}^\mathrm{PRF}_{\F(\kappa)} (2^{w},\allowbreak O(1),\allowbreak 2^w)$, and
unreliability is at most $l(2^{-h+w}+
{\bf InSec}^\mathrm{PRF}_{\F(\kappa)} (2^{w},O(1),2^w))$.
\end{theorem}

\begin{proof}
Insecurity bound is apparent from the fact that if $\F$ were truly random,
then the system would be perfectly secure, because its output is
distributed identically to $\CC$ (simply because the encoder samples from
the channel and independently at random decides which sample to output,
because the random function is never applied more than once to the same input).
Hence, any adversary for the stegosystem would distinguish $\F$ from
random.

The reliability bound per symbol can be demonstrated as
follows.  Assuming that $\F$ is random, the probability that $f$ becomes
1 after $j$ iterations of the inner loop in $\stf.\SE$ (i.e., that
$s_i=s_{i,j}$)
is $(1-2^{-w})^{j-1}2^{-w}$.  If that happens, the probability that $\exists
j'<j$ such that $s_{i,j}=s_{i,j'}$ is at most $(j-1)2^{-h}$.  Summing up
and using standard formulas for geometric series,
we get 
\[ \sum_{j=1}^{\infty} (j-1)2^{-h}\left(1-2^{-w}\right)^{j-1}2^{-w}
= 2^{-h-w}\sum_{j=1}^{\infty}
\left(\left(1-2^{-w}\right)^j\left(\sum_{k=0}^{\infty} (1-2^{-w})^k\right)\right)
< 2^{w-h}. \]
\end{proof}

Note that errors are independent for each symbol, and hence
error-correcting codes over alphabet of size $2^w$ can be used to increase
reliability: one simply encodes $m$ before feeding it to $\SE$.  Observe
that, for a truly random $\F$, if an error occurs in position $i$, the
symbol decoded is uniformly distributed among all elements of
$\bool^w-\{m_i\}$.  Therefore, the stegosystem creates
a $2^{w}$-ary symmetric channel with error
probability $2^{w-h}(1-2^{-w})= 2^{-h}(2^w-1)$ (this comes from more
careful summation in the above proof).  Its capacity is
$w-H[1-2^{-h}(2^w-1), 2^{-h}, 2^{-h}, \dots, 2^{-h}]$
(where $H$ is Shannon
entropy of a distribution) \cite[p. 58]{McE02}.  This is equal
to $w+(2^w-1)2^{-h}\log 2^{-h}+ (1-2^{-h}(2^w-1))\log (1-2^{-h}(2^w-1))$.
Assuming that the error probability $2^{-h}(2^w-1)\le 1/2$ and using
$\log (1-x) \ge -2x$ for $0\le x\le 1/2$, we get
that the capacity of the channel created by the encoder 
is at least $w+2^{-h}(2^w-1)(-h-2)\ge w-(h+2) 2^{-h+w}$.  Thus, as $l$
grows, we can achieve rates close to $w-(h+2)2^{-h+w}$ with near perfect
security and reliability (independent of $h$).

\subsection{Stateless Variants of $\stf$}
\label{section-stf-stateless}
Our stegosystem $\stf$ is stateful because we need $F$ to take $\ctr$ as
input to make sure we never apply the pseudorandom function more than once
to the same input.  This will happen automatically, without the need for
$\ctr$, if the channel $\CC$ has the following property: for any histories
$\his$ and $\his'$ such that $\his$ is the prefix of $\his'$, the supports
of $D_\his$ and $D_{\his'}$ do not intersect.  For instance, when documents
have monotonically increasing sequence numbers or timestamps, no shared
state is needed.

To remove the need for shared state for all channels, we can do the
following.  We remove $\ctr$ as an input to $F$ and instead provide
$\stf.\SE$ with the set $Q$ of all values received so far as answers from
$M$.  We replace the line ``\kwif $\exists j'<j$ s.t. $s_{i,j}=s_{i,j'}$''
with ``\kwif $s_{i,j}\in Q$'' and add the line ``$Q\gets
Q\cup\{s_{i,j}\}$'' before the end of the inner loop.  Now shared state is
no longer needed for security, because we again get fresh coins on each
draw from the channel, even if it collides with a draw made for a previous
hiddentext symbol.  However, reliability suffers, because the larger $l$
is, the more likely a collision will happen.  A careful analysis, omitted
here, shows that unreliability is $l^2 2^{-h+w}$ (plus the insecurity of
the PRF).

Unfortunately, this variant requires the encoder to store the set $Q$
of all the symbols
ever sampled from $\CC$. Thus, while it removes shared state, it requires a
lot of private state.  This storage can be reduced somewhat by use of Bloom
filters~\cite{Blo70} at the expense of introducing potential false collisions
and thus further decreasing reliability. An analysis
utilizing
the bounds of \cite{BM02} (omitted here) shows that using a Bloom filter
with $(h-w-\log l)/\ln 2 $ bits per entry will increase unreliability by
only a factor of 2, while potentially reducing storage significantly
(because the symbols of $\Sigma$ require at least $h$ bits to store
and possibly more if the $D_\his$ is sparse).  

\section{The Stateless Construction $\stl$}

The stateless construction $\stl$ is simply $\stf$ without the
counter and collision detection (and is a generalization to
rate $w$ of the construction that appeared in the extended abstract of
\cite{HLvA02cor}).  Again, we emphasize that the novelty is not in the
construction but in the analysis.  The construction requires a reliability
parameter $k$ to make sure that expected running time of the encoder does
not become infinite due a low-probability event of infinite running time.

\medskip
\noindent
\begin{tabular}{ll}
\begin{minipage}{3.5in}
\begin{tabbing}
{\bf Procedure $\stl.\SE(K,w,k,m,\his)$}:\\
123\=123\=123\=123\=123\=\kill
\>Let $m=m_1\dots m_l$, where $|m_i|=w$
\hspace{1.27in}  \\
\>\kwfor $i\gets 1$ \kwto $l$:\\
\>\> $j\gets 0$\\
\>\> \kwrepeat:\\
\>\>\>$j\gets j+1$\\
\>\>\>$s_{i,j} \gets M(\his)$\\
\>\> \kwuntil $F_K(s_{i,j})=m_i$ \kwor $j=k$\\
\>\> $s_i\gets s_{i,j}$; $\his\gets \his || s_i$\\
\> \kwoutput $s=s_1 s_2 \dots s_l$
\end{tabbing}
\end{minipage}

&

\begin{minipage}{3in}
\begin{tabbing}
123\=123\=123\=\kill
{\bf Procedure $\stl.\SD(K,w,s)$}:\\
\> Let $s=s_1\dots s_l$, where $s_i\in\Sigma$\\
\>\kwfor $i=1$ \kwto $l$\\
\>\>$m_i \gets F_K(s_i)$\\
\> \kwoutput $m = m_1m_2\cdots m_l$\\
\\
\\
\\
\\
\end{tabbing}
\end{minipage}
\end{tabular}

\begin{theorem}
\label{theorem-stateless-security}
The stegosystem $\stl$ has insecurity
\begin{eqnarray*}
\InSec^\mathrm{SS}_{\stl(\kappa, w,
k),\CC} (t, d, l, lw) & \in & 
     O(2^{-h+2w} l^2 + l e^{-k/2^w}) 
   +\InSec^\mathrm{PRF}_{\F(\kappa)} (t+O(1), d+O(1),l2^{w}) \,.
\end{eqnarray*}
\noindent
More precisely,
\begin{eqnarray*}
\lefteqn{\InSec^\mathrm{SS}_{\stl(\kappa, w, 
k),\CC} (t, d, l, lw) <}\hspace{.5in} \\
& &  2^{-h} \left( 
  l(l+1) 2^{2w}
  -  l(l+3) 2^{w} + 2l
  \right)
  +2l \left( 1- \frac{1}{2^{w}} \right)^k
   +\InSec^\mathrm{PRF}_{\F(\kappa)} (t+1,d+O(1), l2^{w}).
\end{eqnarray*}
\end{theorem}

\begin{proof}
The proof of Theorem~\ref{theorem-stateless-security} consists of a hybrid argument. The
first step in the hybrid argument is to replace the stegoencoder $\SE$ with
$\SE_1$, which is the same as $\SE$, 
except that it uses a truly random $G$ instead of
pseudorandom
$F$, which accounts for the term 
$\InSec^\mathrm{PRF}_{\F(\kappa)} (t+O(1), d+O(1),l2^{w})$.
Then, rather than consider directly the statistical difference between
$\CC$ and the output of $\SE_1$ on an $lw$-bit message,
we bound it via a series of steps involving related stegoencoders
(these are not encoders in the sense defined in
Section~\ref{section-definitions}, as they do not have corresponding
decoders; they are simply related procedures that help in the proof).

The encoders $\SE_2$,
$\SE_3$, and $\SE_4$ are specified
in Figure~\ref{figure-encoders}. 
$\SE_2$ is the same
as $\SE_1$, except that it maintains a set $Q$ of all answers received
from $M$ so far. After receiving an answer $s_{i,j}\gets M(\his)$,
it checks if $s_{i,j}\in Q$; if so, it aborts and outputs ``Fail''; else,
it adds $s_{i,j}$ to $Q$.  It also aborts and outputs ``Fail'' if $j$ ever
reaches $k$ during an execution of the inner loop.
$\SE_3$ is the same as $\SE_2$, except that instead of thinking of
random function $G$ as being fixed before hand, it creates $G$ ``on
the fly'' by repeatedly flipping coins to decide the $w$-bit value assigned
to $s_{i,j}$.  Since, like $\SE_2$, it aborts whenever a collision between strings of
covertexts occurs, the function will remain consistent.
Finally, $\SE_4$ is the same as $\SE_3$, except that it never aborts with
failure.

\begin{figure}[t]
\noindent
\begin{tabular}{lll}
\small
\begin{minipage}{2.2in}
\begin{tabbing}
$\SE_2(K,w,k,m_1\dots m_l,\his)$:\\
123\=12\=123\=123\=\kill
 $Q\gets \emptyset$\\
\kwfor $i\gets 1$ \kwto $l$:\\
\> $j\gets 0$\\
\> \kwrepeat:\\
\>\>$j\gets j+1$\\
\>\>$s_{i,j} \gets M(\his)$\\
\>\>\kwif $s_{i,j}\in Q$ \kwor $j=k+1$ \kwthen \\
\>\>\>\kwabort and \kwoutput ''Fail''\\
\>\>$Q \gets Q\cup \{s_{i,j}\}$\\
\\
\> \kwuntil $G(s_{i,j})=m_i$\\
\> $s_i\gets s_{i,j}$; $\his\gets \his || s_i$\\
 \kwoutput $s=s_1 s_2 \dots s_l$\\
\end{tabbing}
\end{minipage}

&

\small
\begin{minipage}{2.2in}
\begin{tabbing}
$\SE_3(K,w,k,m_1\dots m_l,\his)$:\\
3\=12\=123\=123\=123\=\kill
\> $Q\gets \emptyset$\\
\>\kwfor $i\gets 1$ \kwto $l$:\\
\>\> $j\gets 0$\\
\>\> \kwrepeat:\\
\>\>\>$j\gets j+1$\\
\>\>\>$s_{i,j} \gets M(\his)$\\
\>\>\>\kwif $s_{i,j}\in Q$ \kwor $j=k+1$ \kwthen\\
\>\>\>\>\kwabort and \kwoutput ''Fail''\\
\>\>\>$Q \gets Q\cup \{s_{i,j}\}$\\
\>\>\>Pick $c\in_R\{0,1\}^w$\\
\>\> \kwuntil $c=m_i$\\
\>\> $s_i\gets s_{i,j}$; $\his\gets \his || s_i$\\
\> \kwoutput $s=s_1 s_2 \dots s_l$\\
\end{tabbing}
\end{minipage}

&

\small
\begin{minipage}{2.2in}
\begin{tabbing}
$\SE_4(K,w,k,m_1 \dots m_l,\his)$:\\
3\=123\=123\=123\=123\=\kill
\\
\>\kwfor $i\gets 1$ \kwto $l$:\\
\>\> $j\gets 0$\\
\>\> \kwrepeat:\\
\>\>\>$j\gets j+1$\\
\>\>\>$s_{i,j} \gets M(\his)$\\
\\
\\
\\
\>\>\>Pick $c\in_R\{0,1\}^w$\\
\>\> \kwuntil $c=m_i$\\
\>\> $s_i\gets s_{i,j}$; $\his\gets \his || s_i$\\
\> \kwoutput $s=s_1 s_2 \dots s_l$\\
\end{tabbing}
\end{minipage}
\end{tabular}
\caption{``Encoders'' $\SE_2$, $\SE_3$, and $\SE_4$ used in the proof of
Theorem~\ref{theorem-stateless-security}}
\label{figure-encoders}
\end{figure}

In a sequence of lemmas, we
bound the statistical
difference between the outputs of 
$\SE_1$ and $\SE_2$; show
that it is the same as the statistical difference
between the outputs of $\SE_3$ and $\SE_4$; and show
that the outputs of $\SE_2$ and $\SE_3$ are distributed identically.
Finally, observe that $\SE_4$ does nothing more than sample from the
channel and then randomly and obliviously to the sample keep or discard
it.  Hence, its output is distributed identically to the channel.
The details of the proof follow.

For ease of notation, we will denote $2^{-h}$ (the upper bound on the
probability of elements of $D_\his$) by $\maxprob$ and $2^w$ by $R$ for
the rest of this proof.

The following proposition serves as a warm-up for the proof of
Lemma~\ref{lemma-RStwo-RStwoa-multi}, which follows it.
\begin{proposition}\label{prop-RStwo-RStwoa-SD}
  The statistical difference between the output distributions of
  $\SE_1$ and $\SE_2$ for a $\chanRate$-bit hiddentext message
  $m \in\{0,1\}^{\chanRate}$ is at most
  $2\maxprob/(R-1)^2+2\natE^{-\RScountk/R}$.That is, 
  \begin{eqnarray*}
    \sum_{\forall s\in \Sigma} 
      \left|\Pr_{G,M}[\SE_1(K,w,k,m,\his)\to s] -
            \Pr_{G,M}[\SE_2(K,w,k,m,\his)\to s]\right|\\
    <  2\maxprob (R-1)^2
    +2\natE^{-\RScountk/R}\,.
  \end{eqnarray*}
\end{proposition}

\begin{proof}
  Consider the probability that $\SE_2$ outputs ``Fail'' while trying to
  encode some $m \in \{0,1\}^{\chanRate}$.  This happens for one of two
  reasons. First, if after $\RScountk$ attempts to find $s_{i,j}$ such that
  $G(s_{i,j})=m_i$, no
  such $s_{i,j}$ has been drawn.  Second, if the same value is returned twice
by $M$ before $\SE_2$ finds a
  satisfactory $s_{i,j}$; in other
  words, if
  there has been a
  collision between two unsuccessful covertext documents. 

  Let $E_1$ denote the event that one of these two situations has occurred and
  $n_1$ denote the value of $j$  at  which $E_1$
  occurs. Then
  \begin{eqnarray*}
    \Pr[E_1] &\le& \left( \frac{R-1}{R} \right)^2 \maxprob
    +\left( \frac{R-1}{R} \right)^3 2\maxprob +\cdots 
    +\left( \frac{R-1}{R} \right)^{\RScountk-1} (\RScountk-2) \maxprob
    +\left( \frac{R-1}{R} \right)^{\RScountk}\\
    & = & \maxprob \sum_{n_1=2}^{\RScountk-1} 
    \left( \frac{R-1}{R} \right)^{n_1} (n_1-1)
    +\left( \frac{R-1}{R} \right)^{\RScountk}\\
    & < &  \maxprob \left( \frac{R-1}{R} \right)^2 \sum_{n_1=0}^{\infty} 
    \left( \frac{R-1}{R} \right)^{n_1} (n_1+1)
    +\left( \frac{R-1}{R} \right)^{\RScountk}\\
    & = & \maxprob (R-1)^2
    +\left( \frac{R-1}{R} \right)^{\RScountk}\\
    & < & \maxprob (R-1)^2 
    +\natE^{-\RScountk/R}\,.
  \end{eqnarray*}

  Observe that the probability that $\SE_2$ outputs a specific
  document $s$
  which is not ``Fail'' can be only less than the probability that
  $\SE_1$
  outputs the same element. Since the total decrease over all such $s$ is at
  most the probability of failure from above, the total statistical
  difference is at most $2\Pr[E_1]$.
\end{proof}

\begin{lemma}\label{lemma-RStwo-RStwoa-multi}
  The statistical difference between the output of $\SE_1$ and $\SE_2$ 
  when encoding a message $m\in
  \{0,1\}^{l \chanRate}$ is at most
\begin{eqnarray*}
  \maxprob \left( 
  l(l+1) R^2
  -  l(l+3) R + 2l
  \right)
  +2l \left( 1- \frac{1}{R} \right)^k\,.
\end{eqnarray*}
\end{lemma}

\begin{proof}

Proposition~\ref{prop-RStwo-RStwoa-SD} deals with the case $l=1$.  It
remains 
to extend this line of analysis to the general case $l>1$. As in the
proof of Proposition~\ref{prop-RStwo-RStwoa-SD}, let $E_i$ denote the event
that 
$\SE_2$ outputs ``Fail'' while attempting to encode the $i$th block of
$m_i$. Note that $E_i$
grows with $i$ because the set $Q$ grows as more and more blocks are encoded. 
Also, let $n_i$
denote the number of attempts used by $\SE_2$ to encode the $i$th
block. To simplify the analysis, we initially ignore the boundary case of
failure on attempt $n_i=\RScountk$ and treat a failure on this attempt like
all others. Let $E'_i$ denote these events. Then, we have the following
sequence of probabilities. 

Recall that, for $E'_1$,
\[ \Pr[E'_1] < \maxprob (R-1)^2 \,.\]
In the harder case of $E'_2$,
\begin{eqnarray*}
  \Pr[E'_2] & = & \sum_{n_1=1}^{k}
  \Pr[E'_2|n_1 \mathrm{\ draws\ for\ bit\ 1}] 
  \Pr[n_1 \mathrm{\ draws\ for\ bit\ 1}]\\ 
  &\le& \frac{\maxprob}{R} \sum_{n_1=1}^{k}\sum_{n_2=1}^{k} 
  \left( \frac{R-1}{R} \right)^{n_1+n_2-1} (n_1+n_2-1)\\
  & = & \frac{\maxprob}{R} \sum_{n_1=1}^{k} \left( \frac{R-1}{R} \right)^{n_1-1} 
  \left( 
  \sum_{n_2=1}^{k}\left( \frac{R-1}{R} \right)^{n_2} (n_2-1) 
  +n_1 \sum_{n_2=1}^{k}\left( \frac{R-1}{R} \right)^{n_2}
  \right)\\ 
  & < & \frac{\maxprob}{R} \sum_{n_1=1}^{k} \left( \frac{R-1}{R} \right)^{n_1-1} 
  \left( 
  \Pr[E'_1]/\maxprob +n_1 (R-1)
  \right)\\ 
  & < & \frac{\maxprob}{R} 
  \left(
	R \Pr[E'_1]/\maxprob\ + R^2(R-1)
  \right)\\ 
  & = & \maxprob
	\left( (R-1)^2 + R(R-1)
	\right)\\
  & = & \maxprob (2R-1)(R-1)\,.
\end{eqnarray*}
Similarly, for $E'_3$,
\begin{eqnarray*}
  \Pr[E'_3] &\le& \frac{\maxprob}{R^2} \sum_{n_1=1}^{k} \sum_{n_2=1}^{k}
  \sum_{n_3=1}^{k} \left( \frac{R-1}{R} \right)^{n_1+n_2+n_3-2} 
  (n_1+n_2+n_3-1)\\ 
  & = & \frac{\maxprob}{R^2} \sum_{n_1=1}^{k} \left( \frac{R-1}{R} \right)^{n_1-1} 
  \left( R \Pr[E'_2]/\maxprob 
  + n_1 \sum_{n_2=1}^{k} \left( \frac{R-1}{R} \right)^{n_2-1} \sum_{n_3=1}^{k} 
  \left( \frac{R-1}{R} \right)^{n_3} 
  \right)\\ 
  & < & \frac{\maxprob}{R^2} \sum_{n_1=1}^{k} \left( \frac{R-1}{R} \right)^{n_1-1} 
  \left(  
  R \Pr[E'_2]/\maxprob + n_1 R(R-1)
  \right)\\ 
  & < & \frac{\maxprob}{R^2}
	\left(
	R^2 \Pr[E'_2]/\maxprob + R^3 (R-1)
	\right) \\
  & = & \maxprob (3R-1)(R-1)\,.
\end{eqnarray*}
\noindent In general, for $E'_i$, we have the recurrence
\begin{eqnarray*}
  \Pr[E'_i] &\le& \frac{\maxprob}{R^{i-1}} \sum_{n_1=1}^{k}
  \left( \frac{R-1}{R} \right)^{n_1-1} 
  \left(
  R^{i-2} \Pr[E'_2]/\maxprob +n_1 R^{i-2}(R-1)
  \right)\\
  & < & \Pr[E'_{i-1}] + \maxprob R (R-1)\,,
\end{eqnarray*}
\noindent which when solved yields
\begin{eqnarray*}
\Pr[E'_i] & < & \maxprob (iR-1)(R-1)\,.
\end{eqnarray*}

Now summing up the probability of failure for each of the $\chanRate$-bit
blocks of hiddentext gives
\begin{eqnarray*}
  \sum_{i=1}^{l} \Pr[E'_i] 
  & < & \maxprob (R-1) \sum_{i=1}^{l}(iR-1)\\
  & = & \maxprob (R-1) \left( R \sum_{i=1}^{l}i
  -\sum_{i=1}^{l}1 \right)\\
  & = & \maxprob (R-1) \left( \frac{R l (l+1)}{2} -l \right)\\
  & = & \maxprob 
  \left( 
  \left( \frac{R^2}{2} \right) (l+1)l
  - \left( \frac{R}{2} \right) (l+3)l +l
  \right)\,.
\end{eqnarray*}

Next, we compute the probability of the event that the encoding of block
$m_i$ fails because there were $\RScountk$ unsuccessful attempts to find a
string of $n$ covertexts which evaluates to $m_i$ under $G$, given that
no collisions occurred so far. 
Call this event $\hat{E}_i$. Then
\begin{eqnarray*}
  \Pr[\hat{E}_i] & < & \left( \frac{R-1}{R} \right)^k\,:
\end{eqnarray*}

Finally, we compute the total probability of failure which is at most the
sum of the $E'_i$ and $\hat{E}_i$ events.  That is, the probability that
$\SE_2$ outputs ``Fail'' while encoding any of the $l$ $w$-bit
blocks of $m_i$ of $m$ is at most
\begin{eqnarray*}
 \sum_{i=1}^l \Pr[E_i] & < & \sum_{i=1}^l \Pr[E'_i] + \Pr[\hat{E}_i]\\
 & < & \maxprob 
  \left( 
  \left( \frac{R^2}{2} \right) (l+1)l
  - \left( \frac{R}{2} \right) (l+3)l +l
  \right)
 +l \left( \frac{R-1}{R} \right)^k\,.
\end{eqnarray*}
The statistical difference is at most just twice this amount.
\end{proof}

\begin{lemma}\label{lemma-RStwob-RStwoc-SD}
  The statistical difference between the output distributions of $\SE_2$
  and $\SE_3$ for a random function $G$ and hiddentext message
  $m\in\{0,1\}^{l \chanRate}$ is zero.
\end{lemma}

\begin{proof}
  Both $\SE_2$ and $\SE_3$ abort and output ``Fail'' whenever
  the encoding a block $m_i$ fails. This occurs because either: (1) there are
  $\RScountk$ unsuccessful attempts to find  $s_{i,j}$ such that $G(s_{i,j})=m_i$; or  
  (2) the same document is drawn twice, i.e., there is a collision
  between candidate covertext documents. Hence, $\SE_2$ evaluates
  $G$ at most once on each element of $\Sigma$. So, although $\SE_3$ ignores
  $G$ and creates its own random function by flipping coins at each
  evaluation, since no element of $\Sigma$ will be re-assigned a new
  value, the output distributions of $\SE_2$ and $\SE_3$ are identical.
\end{proof}

\begin{lemma}\label{lemma-RStwoc-RStwod-multi}
  The statistical difference between the output distributions of
  $\SE_3$ and  $\SE_4$ is equal to
  the statistical difference between the output distributions
  of $\SE_1$ and $\SE_2$ used to encode the same message.
\end{lemma} 

\begin{proof}
  As Lemma~\ref{lemma-RStwo-RStwoa-multi} shows, the
  probability that $\SE_2$ (and consequently $\SE_3$ by
  Lemma~\ref{lemma-RStwob-RStwoc-SD}) outputs ``Fail''
  is at most 
\begin{eqnarray*}
  \left( 
  \left( \frac{R^2}{2} \right) (l+1)l
  - \left( \frac{R}{2} \right) (l+3)l +l
  \right)
 +l \left( \frac{R-1}{R} \right)^k\,.
\end{eqnarray*}
  Note that $\SE_4$ has no such
  element; the probabilities of each output other that ``Fail''
  can only increase.  Hence, the total statistical difference is twice
 the probability of ``Fail.''
\end{proof}

These three Lemmas, put together, conclude the proof of the Theorem.  We
can save a factor of two in the statistical difference by the following observation.
Half of the statistical difference between the outputs of
$\SE_1$ and $\SE_2$, as well as between the outputs of $\SE_3$ and $\SE_4$,
is due to the probability of ``Fail''.  Because neither $\SE_1$ nor $\SE_4$
output ``Fail,'' the statistical difference between the distributions they
produce is therefore only half of the sum of the statistical differences.
\end{proof}

\begin{theorem}
\label{theorem-stateless-reliability}
The stegosystem $\stl$ has unreliability
\begin{eqnarray*}
\UnRel^\mathrm{SS}_{\stl(\kappa, w,
k),\CC,l} \le l\left(2^w\exp\left[-2^{h-2w-1}\right]+\exp\left[-2^{-w-1}k\right]\right)
   +\InSec^\mathrm{PRF}_{\F(\kappa)} (t, d, l2^{w}) \,,
\end{eqnarray*}
where $t$ and $d$ are the expected running time and description size,
respectively, of the stegoencoder and the stegodecoder combined.
\end{theorem}

\begin{proof}
As usual, we consider unreliability if the encoder is using a truly
random $G$; then, for a pseudorandom $F$, the encoder and decoder will act as a
distinguisher for $F$ (because whether something was encoded correctly can
be easily tested by the decoder), which accounts for the $\InSec^{PRF}$
term.

The stegoencoder fails to encode properly when it cannot find
$s_{i,j}$ such that $G(s_{i,j})=m_i$ after $k$ attempts.  We will
consider separately the case where $G$ is simply unlikely to hit
$m_i$ and
where $G$ is reasonably likely to hit $m_i$, but the samples from the
channel are just unlucky for $k$ times in a row.

To bound the probability of failure in the first case,
fix some channel history $\his$ and $w$-bit message $m$ and consider the
probability over $G$ that
$G(D_\his)$ is so skewed that the weight of $G^{-1}(m)$ in
$D_\his$ is less $c 2^{-w}$ for some constant $c<1$ (note that the expected
weight is $2^{-w}$).  Formally, consider 
$\Pr_G[\Pr_{s\leftarrow D_\his} [G(s)=m]<c 2^{-w}]$.
Let $\Sigma=\{s_1 \dots s_n\}$ be the alphabet,
and let $\Pr_{D_\his}[s_i]=p_i$.  Define the random variable $X_i$ as $X_i=0$ if
$G(s_i)=m$ and $X_i=p_i$ otherwise.  Then the weight of
$G^{-1}(m)$ equals $\Pr_{s\leftarrow D_\his}[G(s)=m]=1-\sum_{i=1}^n X_i$.
Note that the expected value, over $G$, of
$\sum_{i=1}^n X_i$ is $1-2^{-w}$. Using Hoeffding's inequality
(Theorem 2 of \cite{Hoe63}), we obtain 
\begin{eqnarray*}
\Pr_G[1-\sum_{i=1}^n X_i \le c 2^{-w}]& \le & \exp\left[-2(1-c)^2 2^{-2w}/\sum_{i=1}^n
p_i^2\right] \\
& \le & \exp\left[-2(1-c)^2 2^{-2w}/2^{-h} /\sum_{i=1}^n
p_i\right] \\
& = & \exp\left[-2(1-c)^2 2^{h-2w}\right]\,,
\end{eqnarray*}
where the second to last step follows from $p_i \le 2^{-h}$ and the last
step follows from $\sum_{i=1}^n p_i = 1$.  If we now set $c=1/2$ and take
the union bound over all messages $m\in \bool^w$, we get that the
probability
that $G$ is skewed for at least one message is at most
$2^w\exp\left[-2^{h-2w-1}\right]$.

To bound the probability of failure in the second case, assume that
$G(D_\his)$ is not so skewed.  Then
the probability of failure is
\[
(1-c2^{-w})^k\le \exp\left[-c2^{-w}k\right]\,.
\]
The result follows from setting $c=1/2$ and taking the union bound over $l$.
\end{proof}

\section*{Acknowledgments}
We are grateful to Nick Hopper for clarifying related work and to anonymous
referees for their helpful comments.
                                                                                
The authors were supported in part by the National Science Foundation
under Grant No. CCR-0311485.  Scott Russell's work was also facilitated in
part by a National Physical Science Consortium Fellowship and by stipend
support from the National Security Agency.


\appendix

\section{On Using Public $\eps$-Biased Functions}
\label{appendix-eps-biased}
Many stegosystems~\cite{HLvA02cor,vAH04,BC05} (particularly public-key
ones) use the following approach: they encrypt the hiddentext using
encryption that is indistinguishable from random and then use rejection
sampling with a public function $f:\Sigma\to\{0,1\}^w$ to stegoencode the
resulting ciphertext.

For security, $f$ should have small bias on $D_\his$: i.e., for every $c\in
\{0,1\}^w$, $\Pr_{s\in D_\his}[s\in f^{-1} (c)]$ should be close to $2^{-w}$.
It is commonly suggested that a universal hash function with a published seed
(e.g., as part of the public key) be used for $f$.

Assume that the stegosystem has to work with a memoryless channel $\CC$, i.e.,
one for which the distribution $D$ is the same regardless of history.  Let
$E$ be the distribution induced on $\Sigma$ by the following process:
choose a random $c\in\{0,1\}^w$ and then keep choosing $s\in D$ until
$f(s)=c$.  Note that the statistical difference between $D$ and $E$ is
exactly the bias $\eps$ of $f$.  We are interested in the statistical
difference between $D^l$ and $E^l$.

For a universal hash function $f$ that maps a distribution of min-entropy $h$
to $\bool^w$, the bias is roughly $\eps=2^{(-h+w)/2}$.  As shown in
\cite{Rey04}, if $l<1/\eps$ (which is reasonable to assume here),
statistical difference between $D^l$ and $E^l$ is roughly at least
$\sqrt{l}\eps$.

Hence, the approach based on public hash functions results in statistical
insecurity of about $\sqrt{l}2^{(-h+w)/2}$.

\end{document}